\documentclass[journal, comsoc]{IEEEtran}
\usepackage{amsmath,amssymb}
\usepackage{graphicx}
\usepackage{titlesec}
\graphicspath{{figures/}}
\usepackage{cite}
\usepackage{booktabs}
\usepackage{caption}
\usepackage{siunitx}
\usepackage{balance}
\usepackage{url}
\usepackage{tikz}
\usepackage{float}
\usepackage{xcolor}
\usetikzlibrary{calc, positioning, arrows.meta, patterns, shapes}

\usepackage[pagebackref=false, hidelinks]{hyperref} 
\hypersetup{
    colorlinks=true,       
    linkcolor=black,        
    citecolor=black,        
    urlcolor=black,         
    bookmarksnumbered=true 
}
\usepackage{bookmark}      

\usepackage{algorithm}
\usepackage{algpseudocode}
\usepackage{subfig}
\usepackage{etoolbox}
\AtBeginEnvironment{algorithmic}{\footnotesize}

\titleformat{\subsubsection}
  {\normalfont\normalsize\itshape}   
  {\arabic{subsubsection})}         
  {1em}                            
  {}

\title{AdaPTwin: Adaptive Multi-Fidelity Predictive Digital Twin for Proactive Radio Resource Management in Vehicular Networks}

\author{Armin Makvandi,  Md. Zoheb Hassan, ~\IEEEmembership{Member,~IEEE}, Md. Jahangir Hossain, ~\IEEEmembership{Senior Member,~IEEE}%
\thanks{A. Makvandi and M. J. Hossain are with the School of Engineering, The University of British Columbia, Kelowna, BC, Canada (e-mail: armin.makvandi@ubc.ca; jahangir.hossain@ubc.ca). M. Z. Hassan is with the Department of Electrical and Computer Engineering, Universit\'e Laval, Qu\'ebec City, QC, Canada (e-mail: md-zoheb.hassan@gel.ulaval.ca).}%
}

\makeatletter
\def\ps@IEEEtitlepagestyle{%
  \def\@oddhead{\parbox{\textwidth}{\centering \footnotesize This work has been submitted to the IEEE for possible publication. Copyright may be transferred without notice, after which this version may no longer be accessible.}}%
  \def\@evenhead{}%
  \def\@oddfoot{}%
  \def\@evenfoot{}%
}
\makeatother

\begin{document}
\maketitle

\begin{abstract}
The highly dynamic nature of vehicular networks necessitates proactive and site-specific radio resource management (RRM) for safety-critical services. While Network Digital Twins (NDTs) have emerged as a promising enabler, physics-based ray tracing is time-consuming, challenging accurate RRM under strict latency constraints. We propose AdaPTwin, an adaptive multi-fidelity predictive NDT framework for proactive and latency-aware interference management in vehicular networks. Unlike single- and multi-fidelity NDTs with fixed fidelity levels, AdaPTwin selects the NDT fidelity by constrained optimization under a per-slot latency budget. The framework adopts a hierarchical cloud--edge architecture: fidelity selection is performed periodically in the cloud, whereas the proactive RRM loop runs at the edge in every slot. The edge loop predicts the channels between vehicles and roadside units (RSUs) through trajectory forecasting, using a Transformer with continual learning, and look-ahead ray tracing with NVIDIA Sionna over a dynamically updated 3D scene of buildings, roads, and vehicles. A joint RSU beamforming and vehicle--RSU association problem is then solved to maximize the proportionally fair sum rate using a scalable multi-start iterative coordinate descent algorithm. AdaPTwin is assessed in a simulation-based evaluation in which a high-fidelity reference NDT, rather than field measurements, provides the accuracy reference. Comparisons with reactive, single-fidelity, and non-adaptive predictive NDTs and the stochastic 3GPP model show that AdaPTwin adapts to the evaluated traffic densities and urban layouts, where fixed-fidelity DT frameworks either lose accuracy or overrun their deadline. In the evaluated scenarios, AdaPTwin achieves up to 70\% sum-rate gain and 80\% outage probability reduction compared with a non-adaptive predictive multi-fidelity NDT while meeting the per-slot deadline, and it remains within 2.3\% of the sum rate of a per-slot joint-optimization upper bound, whereas per-slot joint optimization requires about 44\,s per decision (about 54 times the latency of AdaPTwin), which renders its scheduling decisions outdated in dynamic vehicular networks.

\end{abstract}


\vspace{-0.3cm}
\section{Introduction}
Effective radio resource management (RRM) is paramount for the realization of next-generation vehicle-to-everything (V2X) networks, which must support safety-critical applications such as autonomous driving \cite{Hussein2022Vehicular}. To guarantee strict quality of service (QoS) for safety-critical services, network operators must dynamically optimize beamforming vectors, power allocation, and user association \cite{Ge2019URLLC}. Accurate and timely acquisition of channel state information (CSI) plays a critical role for the optimality of these RRM tasks.

Of these decisions, analog beam selection and vehicle--RSU association, and with them the coordination of inter-cell interference, are semi-persistent and governed by the large-scale, geometry-dependent channel (path gain, angular support, and blockage), which decorrelates over meters rather than wavelengths and varies slowly. This paper targets this semi-persistent timescale, whereas fast link adaptation remains with the lower layers.  However, acquiring accurate CSI in vehicular networks presents critical challenges due to the highly dynamic nature of these networks. Traditional reactive RRM approaches usually experience outdated CSI measurements in highly dynamic vehicular networks \cite{Gupta2025Proactive}, rendering the RRM decisions inefficient, especially for safety-critical services. Furthermore, in multiple-input multiple-output (MIMO) vehicular communications, the pilot overhead scales linearly with the number of antennas and users, consuming substantial time-frequency resources and severely limiting spectral efficiency \cite{Kocharlakota2021Overhead}. Additionally, the reuse of pilot sequences in dense vehicular networks leads to pilot contamination, where interference from neighboring cells further degrades estimation accuracy \cite{Elhoushy2022PilotContamination}.

To enable proactive RRM while addressing the latency and overhead issues, various proactive channel prediction methods have been proposed, yet they fail to fully resolve the challenges in vehicular networks \cite{Adeogun2025ChannelPrediction}. Kalman filtering \cite{Siebert2026KF}, for instance, typically assumes slowly time-varying shadowing, an assumption that conflicts with the rapid channel fluctuations caused by high mobility in vehicular networks. AI-based  CSI prediction methods \cite{Jiang2022AI, Stenhammar2024AI, Xin2026AI} demand massive amounts of high-quality real-world datasets for training, and the resulting models often struggle to generalize under domain shift (e.g., new areas or traffic patterns) \cite{Akrout2023AIGeneralization}, which is common in vehicular networks.

To overcome these limitations, Network Digital Twins (NDTs) \cite{Masaracchia2022NDT, Lin2023NDT, Villa2024NDT, Zhang2025NDT} have emerged as a promising solution. By incorporating 3D models, continuously updated data, physics-based ray tracing, and what-if analysis, an NDT can replicate the physical network for accurate CSI acquisition required in link-and network-level resource scheduling, effectively reducing pilot overhead. Although high-fidelity, material-aware, and multi-interaction ray-tracing can accurately replicate RF propagation phenomena in 3D environment, its high computational complexity limits its scalability for real-time applications \cite{Zhu2024RayTracingLatency}. This necessitates a predictive, adaptive NDT framework capable of balancing the tradeoff between accuracy and latency
with the awareness of dynamic vehicular network conditions.

Furthermore, the performance of any NDT-optimized RRM algorithm depends on both the accuracy of the NDT and the algorithm’s optimality gap within the NDT domain \cite{Tao2025Provable}. In dynamic vehicular networks, both must be kept small, since low-fidelity NDTs rely on oversimplified propagation models, which can render the NDT-optimized RRM ineffective. Although higher NDT fidelity is desirable for accurate RF propagation modeling, it must remain situation-aware and feasible under strict latency constraints, since the required accuracy depends on instantaneous conditions such as vehicle density, mobility, and environmental complexity.  In particular, fidelity selection should ensure graceful degradation rather than catastrophic failure, even under highly dense traffic conditions. Meanwhile, both fidelity selection and NDT reconfiguration are computationally intensive processes.  Therefore, designing an adaptive NDT framework requires carefully balancing accuracy, update frequency, latency, and computational overhead, while accounting for dynamic mobility conditions.

To address the aforementioned challenges, this paper proposes AdaPTwin, an adaptive multi-fidelity predictive NDT framework for proactive interference management in downlink vehicle-to-infrastructure (V2I) networks. Leveraging edge–cloud collaboration, AdaPTwin jointly optimizes two key aspects: NDT fidelity selection to minimize the channel prediction error of the twin, and proactive RRM to enhance co-channel interference management while enabling spectrum sharing among V2I links. Unlike existing DT-in-the-loop RRM approaches, AdaPTwin jointly optimizes both processes while taking dynamic mobility and environmental conditions  into account, thereby achieving superior interference management under strict latency constraints. The specific contributions of this paper are summarized as follows.

\smallskip \indent $\bullet$ We propose AdaPTwin, a DT-in-the-loop framework for proactive and situation-aware interference management in downlink V2I networks that coordinates multi-RSU beam activation and vehicle--RSU association to maximize the proportionally fair sum rate without over-the-air acquisition of global CSI. To the best of our knowledge, it is the first framework to exploit the NDT fidelity (ray-tracing parameters and vehicle-mesh fidelity) as a decision variable of this RRM problem under an end-to-end decision latency budget.

\smallskip \indent $\bullet$ Since the RRM problem remains NP-hard even for a fixed NDT fidelity, and the error and latency of each fidelity configuration are known only after ray tracing, per-slot joint optimization is practically infeasible. We therefore decompose it into a periodic cloud-side fidelity selection over $420$ configurations and a per-slot edge-side proactive decision-making loop that comprises positioning, Transformer-based trajectory prediction with continual learning, 3D scene update, look-ahead ray tracing with NVIDIA Sionna, and RRM by a multi-start iterative coordinate descent algorithm of polynomial complexity.

\smallskip \indent $\bullet$ We show that, in the evaluated scenario, the decomposition remains within $2.3\%$ of the sum rate  of an oracle per-slot joint-optimization bound with the loss stemming from the fidelity selection rather than the RRM, whereas per-slot joint optimization requires about $44$\,s or more per decision, yielding outdated RRM decisions and a substantially higher outage probability than AdaPTwin. The proposed heuristic RRM also achieves a higher sum rate than Gurobi and cuOPT solver-driven optimizations.

\smallskip \indent $\bullet$ We demonstrate AdaPTwin's effectiveness over state-of-the-art reactive (single- and multi-fidelity), single-fidelity predictive, and non-adaptive multi-fidelity NDTs re-implemented in an identical pipeline, and with the 3GPP model, in realistic dense urban vehicular mobility environment, complemented by ablation, sensitivity, and Monte-Carlo analyses.

\textit{Scope.} The evaluation in this paper is simulation-based. We assume that the geometry and materials of the NDT have been calibrated offline with existing methods, and we use a high-fidelity reference NDT as the accuracy reference. The reported RMSE and blockage figures therefore quantify intra-twin fidelity, whereas the computation latencies are measured on GPU hardware and the trajectory predictor is also evaluated on real-world recordings (Section~VIII).

The rest of this paper is organized as follows. In Section II, related works are discussed. In Section III, the system model and problem formulation are presented. Section IV presents the workflow of AdaPTwin with cloud-edge collaboration architecture. 
Section V details the building blocks of the AdaPTwin pipeline, explaining offline and online components, and Section VI develops its algorithms. Section VII presents the simulation results, Section VIII discusses the scope and limitations of the study, and Section IX concludes the paper.

\vspace{-0.3cm}

\section{Related Work}
NDTs for vehicular networks can be broadly categorized into offline, reactive, and predictive frameworks. Offline NDTs rely on precomputed radio maps generated via extensive ray tracing. In our previous work \cite{Makvandi2025Offline}, we utilized such offline maps to extract path gain values for online SINR computation. However, this study revealed that static maps fail to capture the dynamic environmental changes inherent in vehicular networks, necessitating online ray tracing. Emulation-based twins such as Colosseum \cite{Villa2024NDT} reproduce channels on RF hardware but cannot re-synthesize propagation for new geometries, and VaN3Twin \cite{Gasco2025VaN3Twin} couples SUMO, ns-3, and ray tracing in a full-stack twin, but it does not predict future network states or adapt its ray-tracing fidelity. Neither addresses the online latency--accuracy tradeoff of a predictive twin that must close an RRM loop within one slot.

Reactive NDTs attempt to address this by performing ray tracing for the current network state. Demir \textit{et al}. \cite{Demir2023Reactive} proposed a reactive NDT to improve Quality-of-Service (QoS), yet their approach does not account for the computational cost of ray tracing, which hinders real-time performance. To manage this computational load, Salehi \textit{et al}. \cite{Salehi2024Multiverse} introduced a multi-fidelity reactive NDT called Multiverse that selects ray tracing fidelity based on available resources.  However, the high processing latency makes such reactive NDT pipeline unsuitable for highly dynamic vehicular networks where the channel changes faster than the simulation can update. Ray tracing is an established engine for site-specific beamforming studies and datasets \cite{DegliEsposti2014RT, Alkhateeb2019DeepMIMO}, and differentiable solvers allow its calibration against measurements \cite{Hoydis2024LearningRadio}. Multi-fidelity simulation is mature in computational science \cite{Peherstorfer2018Multifidelity}, and the accuracy--runtime tradeoff of ray launching has been quantified \cite{Zhu2024RayTracingLatency}. However, existing wireless twins \cite{Salehi2024Multiverse, Pegurri2026Proactive} expose fidelity only as resource-triggered or predefined levels, rather than selecting it by constrained optimization against a measured latency budget. Luo \textit{et al}. \cite{Luo2025TwinFidelity} quantify the impact of twinning fidelity on DT-aided CSI feedback, but they analyze the fidelity rather than select it online under a decision deadline.

Consequently, predictive NDTs have emerged to predict future network states and enable look-ahead ray tracing. Elloumi \textit{et al}. \cite{Elloumi2025PRISM} proposed PRISM DT, which leverages Long-Short-Term-Memory (LSTM)-based vehicle trajectory prediction to forecast channel gains for proactive V2I clustering and power allocation. While pioneering, PRISM DT did not incorporate the electromagnetic impact of 3D vehicle bodies, a critical factor for accurate radio propagation. Furthermore, they did not consider the important impact of simulation fidelity on accuracy and latency. In \cite{Makvandi2026AIRTwin}, we proposed AIRTwin, leveraging Transformer-based vehicle trajectory prediction and ray tracing for proactive radio map prediction in vehicular networks. We considered 3D modeling of vehicles, in addition to the buildings and roads, to account for the effect of all objects on wave propagation. However, AIRTwin was a single-fidelity NDT without considering an adaptive balance between accuracy and latency. Pegurri \textit{et al}. \cite{Pegurri2026Proactive} proposed a predictive multi-fidelity NDT for vehicular networks. They used a baseline constant-velocity model and standard Kalman filter for vehicle trajectory prediction and predefined several levels of fidelity for ray tracing. However, their work primarily investigates the impact of fidelity on latency without providing an adaptive mechanism for fidelity selection. Data-driven methods predict CSI \cite{Jiang2022AI, Stenhammar2024AI, Xin2026AI}, beams and blockage \cite{Alrabeiah2020Sub6, Charan2021Vision}, or position-aided beams \cite{Va2019Position}. They are tied to the deployment site, degrade under domain shift \cite{Akrout2023AIGeneralization}, and lack the explicit geometric consistency (e.g., blockage by a specific bus) that interference-aware beam coordination requires.

\begin{table*}[!t]
\caption{Comparison of the Proposed AdaPTwin with Existing Frameworks}
\label{tab:comparison}
\centering
\footnotesize
\renewcommand{\arraystretch}{1.2}
\begin{tabular}{l c c c c l}
\toprule
\textbf{Reference} & \textbf{NDT Approach} & \textbf{Trajectory Prediction} & \textbf{Fidelity} & \textbf{3D Vehicles} & \textbf{RRM} \\
\midrule
Makvandi \textit{et al}. \cite{Makvandi2025Offline} & Offline & $-$ & Single & $-$ & Association + Beam Selection \\
Demir \textit{et al}. \cite{Demir2023Reactive} & Reactive & $-$ & Single & Box (obstacle only) & Tx Power + Directivity + RIS \\
Multiverse \cite{Salehi2024Multiverse} & Reactive & $-$ & Multi-Fidelity & \checkmark & Beam selection \\
VaN3Twin \cite{Gasco2025VaN3Twin} & Reactive & $-$ & Single & \checkmark & Full-stack V2X without RRM \\
PRISM DT \cite{Elloumi2025PRISM} & Predictive & LSTM & Single & $-$ & Clustering + Power Allocation \\
AIRTwin \cite{Makvandi2026AIRTwin} & Predictive & Transformer & Single & \checkmark & $-$ \\
Pegurri \textit{et al}. \cite{Pegurri2026Proactive} & Predictive & Kalman Filter & Predefined Multi-Fidelity & \checkmark & $-$ \\
\textbf{AdaPTwin} & \textbf{Predictive} & \textbf{Transformer} & \textbf{Adaptive Multi-Fidelity} & \textbf{\checkmark} & \textbf{Association + Beam Selection} \\
\bottomrule
\end{tabular}
\end{table*}

\textit{Positioning of this work.} Different from the state-of-the-art  literature, AdaPTwin  enables proactive RRM in vehicular networks using a predictive NDT with adaptive fidelity levels (cf. Table I). Its key distinguishing element is the cloud-side periodic selection of the NDT fidelity with situation awareness (e.g., vehicular network density and 3D environment) such that the edge-side proactive decision loop completes  within the given slot. AdaPTwin extends our prior AIRTwin framework by adding dynamic fidelity selection, the two-timescale decomposition, and latency-budget-aware RRM.

\vspace{-0.2cm}

\section{System Model and Problem Formulation}

\subsection{System Overview and Cloud-Edge Architecture}
We consider a downlink vehicular network supported by an NDT integrated within a hierarchical 6G cloud-edge architecture. The network consists of a set of RSUs denoted by $\mathcal{B} = \{1, \dots, B\}$ and a set of vehicular users (UEs) denoted by $\mathcal{U} = \{1, \dots, U\}$. Each RSU $b \in \mathcal{B}$ is equipped with $N_{tx}$ antennas and utilizes a predefined codebook of transmit beams $\mathcal{W} = \{1, \dots, W\}$.

To balance the conflicting requirements of high-fidelity channel modeling and real-time responsiveness, the NDT framework is distributed across two computational tiers as described below.
\begin{itemize}
    \item \textbf{Cloud Tier:} Hosted on centralized servers with substantial GPU resources. It functions as the \textit{training and calibration engine}, responsible for running a very high-fidelity NDT and optimizing fidelity selection policies. Due to computational capability disparities between edge and cloud, we define a latency scaling factor $\kappa$ (edge-to-cloud latency ratio) to estimate Edge execution times based on cloud measurements.
    \item \textbf{Edge Tier:} Hosted on Multi-access Edge Computing (MEC) servers co-located with RSUs. It functions as the \textit{inference engine}, executing real-time operations including vehicle positioning and trajectory prediction, ray-tracing, and proactive RRM. The computed decisions are transferred to the physical network over feedback links.
\end{itemize}

\vspace{-0.3cm}

\subsection{Channel Model and SINR Formulation}
Let $\mathbf{h}_{u,b} \in \mathbb{C}^{N_{tx} \times 1}$ denote the channel vector from RSU $b$ to user $u$ predicted by NDT, and let $\mathbf{f}_w \in \mathbb{C}^{N_{tx} \times 1}$ denote the precoding vector corresponding to beam $w \in \mathcal{W}$.

Let $\mathbf{w} = \{w_1, \dots, w_B\}$ denote a global network beam state, where $w_b$ is the index of active beam at RSU $b$. The Signal-to-Interference-plus-Noise Ratio (SINR) for user $u$ served by RSU $b$ under beam state $\mathbf{w}$ is given by
\begin{equation}
    \gamma_{u,b,\mathbf{w}} = \frac{P_{tx} \left| \mathbf{h}_{u,b}^H \mathbf{f}_{w_b} \right|^2}{\sigma^2 + \sum_{j \in \mathcal{B} \setminus \{b\}} P_{tx} \left| \mathbf{h}_{u,j}^H \mathbf{f}_{w_j} \right|^2},
\end{equation}
where $P_{tx}$ is the transmission power, $\sigma^2$ is the thermal noise power, and $(\cdot)^H$ denotes the Hermitian transpose. The denominator explicitly calculates the inter-cell interference generated by the specific beams $\{w_j\}_{j \neq b}$ active in the current candidate configuration. The vehicles associated with the single active beam of RSU $b$, collected in $\mathcal{U}_b$, are served on orthogonal time--frequency resources by a link-layer scheduler per RSU. Hence, $\gamma_{u,b,\mathbf{w}}$ does not depend on the resource share $\rho_{u,b}$, the throughput is $\rho_{u,b} B_W \log_2(1+\gamma_{u,b,\mathbf{w}})$, and decisions are taken on the per-user spectral efficiency $\log_2(1+\gamma_{u,b,\mathbf{w}})$, whose sum over the served vehicles is reported as the sum rate in Section VII. With equal shares $\rho_{u,b} = 1/|\mathcal{U}_b|$, the proportional-fairness objective would gain the load-balancing term $-\sum_b |\mathcal{U}_b| \ln |\mathcal{U}_b|$. AdaPTwin and all NDT baselines use the same max-SINR association of Algorithm 3 without this term and are scored with the same metric; the reported sum rate is thus an aggregate spectral efficiency on the allocated resources rather than the delivered cell throughput.

\vspace{-0.3cm}

\subsection{Decision Variables}
Let $\mathcal{C} = \{C_1, \dots, C_K\}$ denote the discrete candidate set of NDT fidelity configurations, where each configuration $C_k \in \mathcal{C}$ is characterized by an estimated channel prediction error $\mathcal{E}(C_k)$ (e.g., path gain RMSE relative to the high-fidelity reference NDT). We define the following binary decision variables for the joint optimization problem:
\begin{enumerate}
    \item \textit{Service Assignment:} $x_{u,b,w} \in \{0,1\}$, where $x_{u,b,w}=1$ if user $u$ is served by RSU $b$ specifically via beam $w$.
    \item \textit{Beam Activation:} $y_{b,w} \in \{0,1\}$, where $y_{b,w}=1$ if RSU $b$ activates beam $w$.
    \item \textit{Fidelity Configuration:} $z_k \in \{0,1\}$, where $z_k=1$ if the NDT adopts parameter configuration $C_k \in \mathcal{C}$.
\end{enumerate}

\vspace{-0.3cm}

\subsection{Joint Optimization Problem}

Our objective is to maximize the proportionally fair network sum rate and minimize the NDT channel prediction error by jointly optimizing vehicle-RSU association, active beam selection, and NDT-fidelity selection, subject to strict latency and communication QoS constraints. Towards this objective, mathematically, we formulate an optimization problem as follows

\begin{subequations} \label{eq:joint_prob}
\begin{align}
\mathcal{P}_{\text{joint}}: \max_{\mathbf{X}, \mathbf{Y}, \mathbf{Z}} \quad & \sum_{u \in \mathcal{U}} \sum_{b \in \mathcal{B}} \sum_{w \in \mathcal{W}} x_{u,b,w} \ln(\log_2(1 + \gamma_{u,b,w})) \nonumber \\
    & - \lambda \sum_{C_k \in \mathcal{C}} z_k \mathcal{E}(C_k) \label{eq:obj} \\
    \text{s.t.} \quad & \sum_{b \in \mathcal{B}} \sum_{w \in \mathcal{W}} x_{u,b,w} \leq 1, \quad \forall u \in \mathcal{U}, \label{eq:c1} \\
    & \sum_{w \in \mathcal{W}} y_{b,w} = 1, \quad \forall b \in \mathcal{B}, \label{eq:c2} \\
    & \sum_{u \in \mathcal{U}} x_{u,b,w} \leq L \cdot y_{b,w}, \quad \forall b \in \mathcal{B}, \forall w \in \mathcal{W}, \label{eq:load} \\
    & x_{u,b,w} = 0, \quad \forall (u,b,w) : \gamma_{u,b,w} < \gamma_{min}, \label{eq:c_sinr} \\
    & \sum_{C_k \in \mathcal{C}} z_k = 1, \label{eq:c4} \\
    & \sum_{C_k \in \mathcal{C}} z_k \tau_{edge}(C_k) + T_{G} \leq T_{H}. \label{eq:c5}
\end{align}
\end{subequations}
In the objective function \eqref{eq:obj}, the first term maximizes the network proportional fairness, while the second term acts as a penalty to minimize the NDT channel prediction error. Specifically, $\mathcal{E}(C_k)$ denotes the estimated channel prediction error (e.g., the path gain RMSE) associated with adopting the NDT fidelity configuration $C_k$, and $\lambda \geq 0$ is a weighting parameter that controls the trade-off between fairness maximization and error minimization. The penalty is measured at each cloud calibration round in terms of the path gain RMSE with respect to the high-fidelity reference NDT $C_{GT}$, defined as\refstepcounter{equation}
\[
\widehat{\mathcal{E}}(C_k) = \Big(\frac{1}{|\mathcal{U}||\mathcal{B}|}\sum_{u,b}\big([G_{u,b}^{C_k}]_{\mathrm{dB}} - [G_{u,b}^{C_{GT}}]_{\mathrm{dB}}\big)^{2}\Big)^{1/2}, \tag{\theequation}\label{eq:err}
\]
where $G_{u,b}^{C} = \sum_{n=1}^{N_{tx}} |\sum_{\ell} a_{u,b}^{(\ell,n)}(C)\, e^{-j2\pi f_c \tau_{u,b}^{(\ell)}(C)}|^2$ is the narrowband path gain with path amplitudes $a$ and delays $\tau$. Once $\mathbf{Z}^\ast$ is fixed, such a penalty is a constant for the edge, so $\lambda$ affects only the DT fidelity selection. We operate in the lexicographic limit $\lambda \to \infty$, allowing the most accurate latency-feasible DT-fidelity configuration to be selected by  Algorithm~2. 

Regarding the constraints, Constraint \eqref{eq:c1} ensures each vehicle is served by only one RSU. Constraint \eqref{eq:c2} enforces the analog beamforming constraint, requiring each RSU to activate exactly one beam. Constraint \eqref{eq:load} couples the service variables to the beam activation variables, ensuring that users are only assigned to active beams and that the number of users per beam does not exceed the load limit $L$. Constraint \eqref{eq:c_sinr} prunes infeasible links that do not meet the minimum SINR threshold $\gamma_{min}$. Constraint \eqref{eq:c4} ensures that exactly one fidelity configuration is selected from the available set $\mathcal{C}$. Finally, Constraint \eqref{eq:c5} is the per-slot deadline constraint: it ensures that the estimated end-to-end edge latency $\tau_{edge}(C_k)$ of the selected configuration (cf. \eqref{eq:budget}), plus a safety guard time $T_{G}$, does not exceed the prediction horizon $T_{H}$.

\textit{Remark 1:} While $\mathcal{P}_{\text{joint}}$ formulates the overarching joint optimization, solving it monolithically in every slot is computationally prohibitive, as it would require predicting channels, evaluating prediction errors $\mathcal{E}(C_k)$, and computing candidate rates across all possibilities in $\mathcal{C}$. AdaPTwin thus decouples $\mathcal{P}_{\text{joint}}$ into a two-timescale hierarchical optimization: the cloud periodically determines the optimal DT fidelity $\mathbf{Z}^*$ (i.e., $C^*$) over a macroscopic interval $T_{\text{update}}$ (e.g., 10\,min) based on prediction accuracy and latency feasibility, and the edge predicts the channel and optimizes real-time beam activation $\mathbf{Y}^*$ and user association $\mathbf{X}^*$ using the DT fidelity configuration $C^*$ in every slot.  

\textit{Remark 2:} The decomposition results in an optimality loss since $C^\ast$ is held fixed over $T_{\text{update}}$ and selected via $\widehat{\mathcal{E}}(C_k)$ instead of the rate term, and the edge RRM is heuristic. It is justified because $C^\ast$ depends on macroscopic statistics that drift over minutes, whereas $\mathbf{X}^\ast$ and $\mathbf{Y}^\ast$ depend on the instantaneous geometry. Section VII-F quantifies the resulting optimality loss.

\vspace{-0.3cm}
\subsection{Complexity Analysis}
The joint optimization problem $\mathcal{P}_{\text{joint}}$ is NP-hard.
Consider a simplified instance where fidelity $\mathbf{Z}$ is fixed. The remaining problem is a discrete beam-selection and association problem whose objective is coupled across RSUs through inter-cell interference and whose search space grows as $W^B$; interference-coupled discrete resource allocation problems of this type are NP-hard in general \cite{Luo2008DSM}. Conversely, the fidelity selection alone is a one-of-$K$ choice that would be trivial if $\mathcal{E}(C_k)$ and $\tau_{edge}(C_k)$ were known; however, these coefficients become available only after ray tracing every candidate, which makes a per-slot solution of $\mathcal{P}_{\text{joint}}$ impractical (Remark~1). Since it contains an NP-hard subproblem, $\mathcal{P}_{\text{joint}}$ is an NP-hard Mixed-Integer Non-Linear Programming (MINLP) problem.

\vspace{-0.3cm}

\section{AdaPTwin Framework}
\subsection{Overview of the Framework}
\begin{figure}[!t] 
  \centering
  \includegraphics[width=\columnwidth,keepaspectratio]{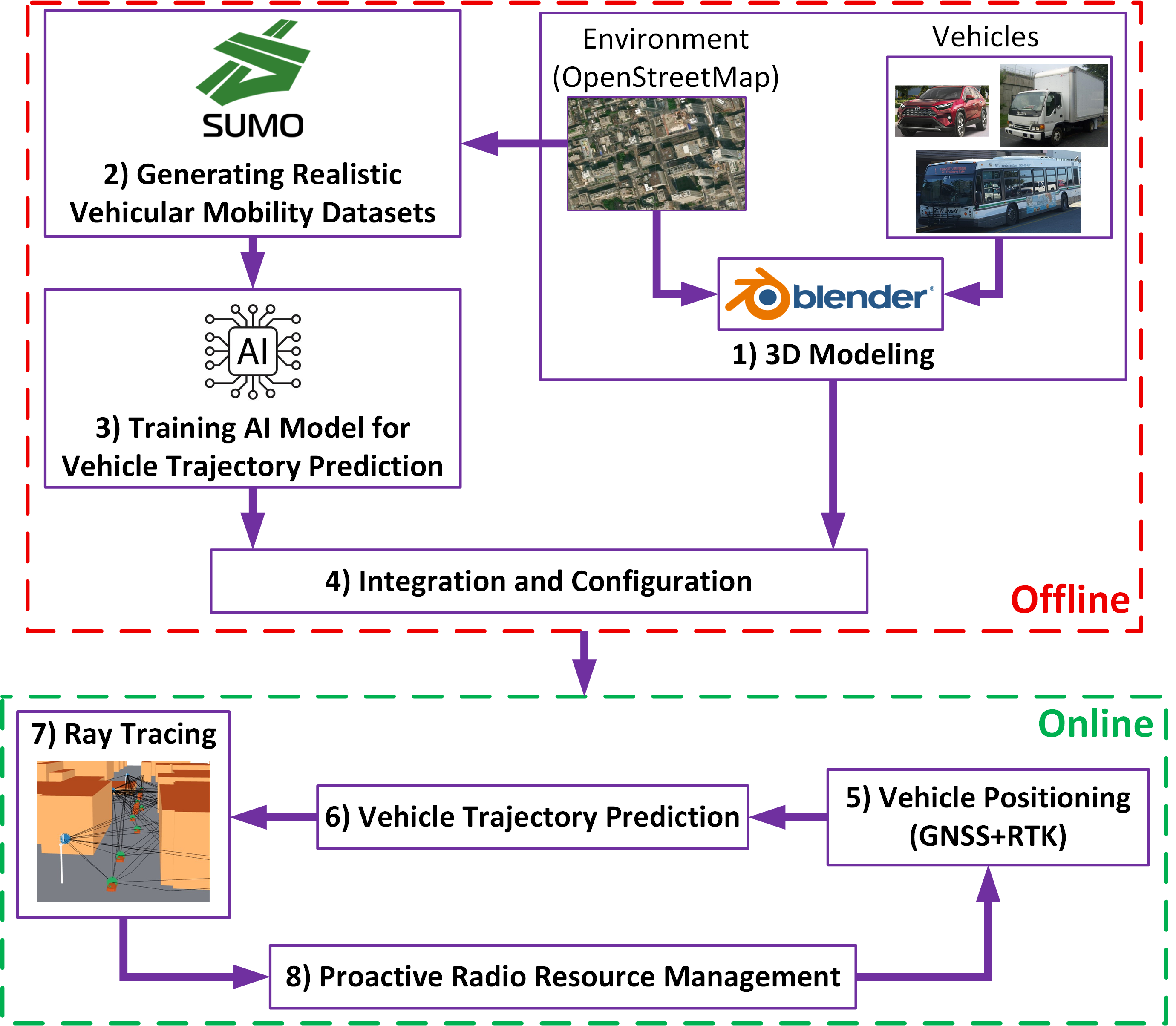}
  \caption{Proposed network digital twin (AdaPTwin) for proactive radio resource management (RRM) in vehicular networks.}
  \label{fig:Twin}
\end{figure}

AdaPTwin consists of offline and online components. The workflow of AdaPTwin is as follows. At the offline component, first, 3D models of the environment and vehicles are generated (1). Then, a realistic vehicular mobility dataset is generated using Simulation of Urban Mobility (SUMO) software \cite{Lopez2018SUMO} (2). An AI model is trained using the generated vehicular mobility dataset for vehicle trajectory prediction (3). At the final step of the offline component (4), the coordinate of the 3D environment is matched with the trajectory dataset. Furthermore, the final 3D models of the environment and vehicles are imported to the radio map prediction engine. In addition, antennas are configured, and RSUs are placed at desired positions. The next steps are included in the online component. The current position of vehicles is imported (5), where the Global Navigation Satellite System (GNSS) and real-time kinematic (RTK) can be used for precise positioning in a practical deployment of AdaPTwin. The next position of vehicles is predicted using the online vehicle trajectory prediction engine (6). The position and heading of 3D vehicles and their corresponding receivers are updated based on predicted trajectories. The next step is ray tracing (7), where online ray tracing is performed using the NVIDIA Sionna library, and channel impulse responses (CIRs) are generated. In this way, CIR is predicted ahead of time, enabling proactive RRM (8) for highly dynamic vehicular networks. The overall structure of the AdaPTwin is shown in Fig. $\,$\ref{fig:Twin}.
\setlength{\textfloatsep}{0pt}

\vspace{-0.3cm}

\subsection{Proposed Cloud-Edge Decomposition}
\begin{figure*}[!t]
  \centering
  \includegraphics[width=\textwidth,height=0.48\textheight,keepaspectratio]{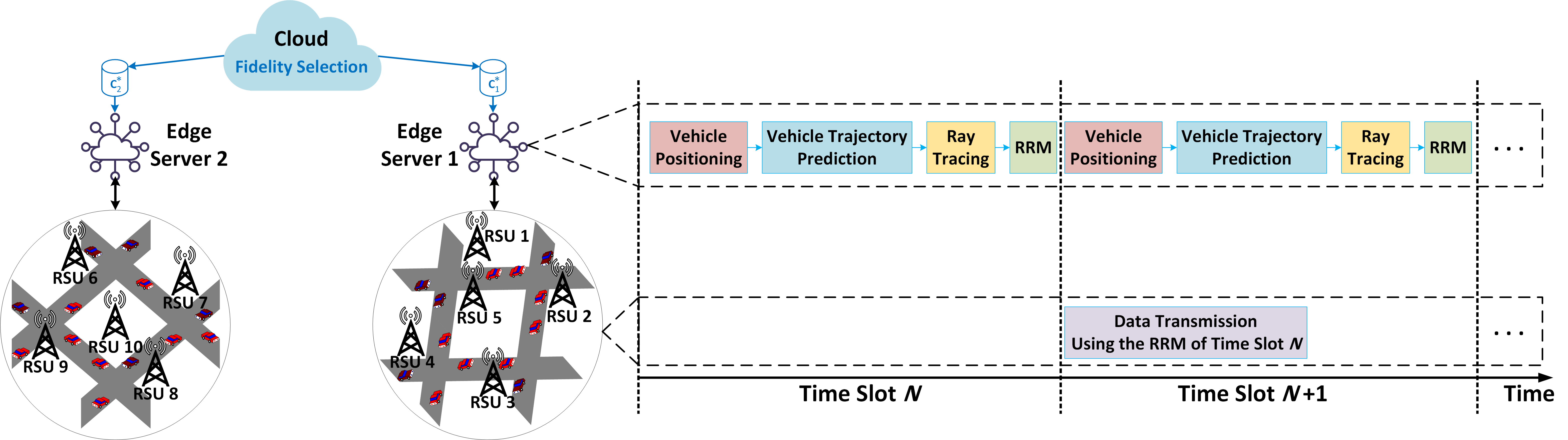}
  \caption{Cloud-Edge based adaptive multi-fidelity framework.}
  \label{fig:cloudedge}
\end{figure*}

As illustrated in Fig. \ref{fig:cloudedge} and Algorithm \ref{alg:AdaPTwin_framework}, AdaPTwin relies on a hierarchical cloud-edge architecture to decouple heavy computations from real-time operations. The centralized cloud tier executes Phase 1 less frequently (e.g., every 10 minutes) to benchmark simulation parameters and select an optimal fidelity policy. Fidelity selection at the cloud yields the most accurate channel prediction that still meets the per-slot deadline. Once deployed to the edge servers, this policy guides Phase 2, which executes in every slot of duration $T_H$. Within Time Slot $N$, the edge pipeline performs vehicle positioning, trajectory prediction, ray tracing, and proactive radio resource management, finalizing the decisions required for actual data transmission in Time Slot $(N+1)$.
\begin{algorithm}[!t]
\caption{AdaPTwin Framework Operation}
\label{alg:AdaPTwin_framework}
\begin{algorithmic}[1]
\Require 3D Environment $\mathcal{S}_{\text{3D}}$, Historical Trajectory Data $\mathcal{D}_{hist}$, Update Interval $T_{update}$
\Ensure Per-Slot Deadline at the Edge, Minimum RMSE, Beam Configuration $\mathbf{w}^*$, User Association $\mathbf{X}^*$

\Statex \textbf{Offline Initialization (Cloud)}
\State Generate 3D digital twin of $\mathcal{S}_{\text{3D}}$ (static map + vehicle models)
\State Train Transformer-based trajectory predictor $\mathcal{M}_{\theta}$ using $\mathcal{D}_{hist}$
\State Deploy $\mathcal{M}_{\theta}$ to Edge Server
\State Initialize fidelity configuration $C^* \gets C_{default}$
\Statex \textbf{Online Operation}
\While{Network Active}
    \Statex \quad \textit{--- Phase 1: (Cloud Tier): Periodic Fidelity Calibration ---}
    \If{$t \mod T_{update} = 0$}
        \State Receive current network context from Edge
        \State $C^* \gets$ \Call{FidelitySelection}{Context} \Comment{See \textbf{Algorithm \ref{alg:fidelity}}}
        \State Send optimized configuration $C^*$ to Edge
    \EndIf
    \Statex \quad \textit{--- Phase 2 (Edge Tier): Proactive RRM (Every Slot) ---}
    \State Obtain vehicles positions $\mathbf{P}_{t-\tau}$ via GNSS-RTK
    \State Predict future positions: $\hat{\mathbf{P}}_{t} \gets \mathcal{M}_{\theta}(\mathbf{P}_{t-\tau})$
    \State Update 3D scene geometry with $\hat{\mathbf{P}}_{t}$
    \State Compute CIR $\mathbf{H}$ using Sionna RT with config $C^*$
    \State $(\mathbf{w}^*, \mathbf{X}^*) \gets$ \Call{HeuristicRRM}{$\mathbf{H}, \mathcal{W}$} \Comment{See \textbf{Algorithm \ref{alg:rrm}}}
    \State Execute downlink transmission using $\mathbf{w}^*$ and $\mathbf{X}^*$
\EndWhile
\end{algorithmic}
\end{algorithm}
\setlength{\textfloatsep}{0pt} 

\vspace{-0.2cm}

\section{Detailed Design of AdaPTwin Building Blocks}

\subsection{3D Virtual Environment Modeling}
Radio propagation is affected by the environment and objects. 3D buildings, roads, and vehicles are considered to represent the real-world environment and objects. OpenStreetMap, an open geographic database, is used to import information about buildings and roads of the desired environment to Blender \cite{Blender}, an open-source 3D computer graphics software, to create the 3D representation of the environment.

3D models of vehicles are also imported into Blender. Detailed 3D models of vehicles are used in the cloud for high-fidelity reference NDT and benchmarking. Using these models maximizes the accuracy of ray-tracing but increases the latency significantly. Since ray-tracing complexity and accuracy depend on the fidelity of 3D vehicle models, simplified 3D vehicle models are exploited for real-time edge processing. Notably, detailed 3D models of vehicles are simplified into several levels of fidelity using low-poly meshes in Blender. The selection of appropriate fidelity of 3D models is performed by Phase 1 optimization in Algorithm \ref{alg:AdaPTwin_framework}.

\vspace{-0.3cm}

\subsection{AI-Assisted Vehicle Trajectory Prediction}

\subsubsection{\textbf{Datasets}}

Simulation of Urban Mobility (SUMO) software is used to generate realistic vehicular mobility datasets. The desired environment is imported from OpenStreetMap into SUMO, and different types of vehicles and the number of each are chosen. In this way, various scenarios of vehicular mobility are generated to test the framework. To develop an AI model for vehicle trajectory prediction that generalizes across scenarios, a complex urban area is selected, and a large and diverse dataset is generated. 

\subsubsection{\textbf{Training AI Model (Offline)}}

To enable proactive RRM in highly dynamic vehicular networks, AdaPTwin predicts future vehicle states to perform ray-tracing and beam selection before the physical transmission occurs. The generated vehicular mobility dataset using SUMO is used to train an AI model for vehicle trajectory prediction. We propose a hybrid framework consisting of an offline-trained Transformer-based backbone for learning temporal motion priors, and a continual-learning mechanism for handling domain shifts online.

The objective of the trajectory prediction module is to map a history of observed states to a target position at the end of a prediction horizon. Let $\mathcal{S}_t = \{\mathbf{s}_{t-K+1}, \dots, \mathbf{s}_t\}$ denote the sequence of historical states over a window $K$, where $\mathbf{s}_k$ represents the state vector at time step $k$. Rather than predicting every intermediate step, the model directly regresses the final future position $\hat{\mathbf{p}}_{t+H}$ at horizon $H$.

\textbf{Input Representation and Preprocessing:}
To generalize across different locations and driving directions, we employ an ego-centric coordinate transformation. For a prediction instance at time $t$, the coordinate system is translated to the vehicle's current position $\mathbf{p}_t$ and rotated such that the vehicle's heading aligns with the positive x-axis. The rotation matrix $\mathbf{R}(\theta_t)$ is defined by the vehicle's instantaneous heading $\theta_t$. The transformed input features $\mathbf{s}'_k$ for a time step $k \in [t-K+1, t]$ are given by
\begin{equation}
    \mathbf{p}'_k = \mathbf{R}(\theta_t)^\top (\mathbf{p}_k - \mathbf{p}_t), \quad \mathbf{v}'_k = \mathbf{R}(\theta_t)^\top \mathbf{v}_k,
\end{equation}
where $\mathbf{p}_k$ and $\mathbf{v}_k$ are the global position and velocity vectors. To capture the vehicle's orientation dynamics, we explicitly append the historical global heading angles $\theta_k$ directly to the transformed kinematic features. The resulting feature vector for each time step is $\mathbf{x}_k = [\mathbf{p}'_k, \mathbf{v}'_k, \theta_k] \in \mathbb{R}^{5}$. Finally, the input sequence is normalized using a standard scaler fitted on the training corpus.

\textbf{Transformer-Based Temporal Encoding:}
The core of the offline engine is a Transformer Encoder architecture designed to capture long-range temporal dependencies. The input sequence $\mathbf{X}' \in \mathbb{R}^{K \times F}$ is first projected to a latent dimension $d_{model}$ via a linear embedding layer. A fixed sinusoidal positional encoding is added to the projected features to retain sequential order without introducing additional trainable parameters.

The sequence is processed by $N_L$ stacked Transformer encoder blocks. Each block consists of a Multi-Head Self-Attention (MHSA) mechanism and a Feed-Forward Network (FFN). The attention mechanism computes \cite{Vaswani2017Attention}
\begin{equation}
    \text{Attention}(\mathbf{Q}, \mathbf{K}, \mathbf{V}) = \text{softmax}\left(\frac{\mathbf{Q}\mathbf{K}^\top}{\sqrt{d_k}}\right)\mathbf{V},
\end{equation}
where $\mathbf{Q}, \mathbf{K}, \mathbf{V}$ are query, key, and value matrices derived from the input. Layer normalization and residual connections are applied after the MHSA and FFN sub-layers. The temporal feature vector is extracted from the final time step of the encoder output to form a single context vector $\mathbf{h}_{ctx}$. This context vector is passed to a Multi-Layer Perceptron (MLP) decoder to regress the final predicted relative displacement $\Delta \hat{\mathbf{p}}_{t+H}$.

\textbf{Hyperparameter Optimization:}
To maximize the offline model's accuracy, we employ the Optuna framework \cite{Akiba2019Optuna} with the Tree-structured Parzen Estimator (TPE) sampler. We jointly optimize the model depth $N_L$, latent dimension $d_{model}$, number of attention heads, learning rate, and regularization factors. The model is trained to minimize the Mean Squared Error (MSE) loss, which strongly penalizes larger prediction deviations to ensure tighter alignment with the ground truth
\begin{equation}
\mathcal{L} = \frac{1}{N} \sum_{i=1}^{N} || \mathbf{p}_{t+H}^{(i)} - \hat{\mathbf{p}}_{t+H}^{(i)} ||^2_2,
\end{equation}
where $||\cdot||_2$ denotes the $L_2$ norm representing the Euclidean distance between the actual and predicted relative displacement at the end of the horizon.

\textbf{Evaluation Metric:}
To quantify the prediction accuracy, we utilize the Final Displacement Error (FDE), which measures the Euclidean distance between the predicted position and the ground truth position at the final time step of the prediction horizon $H$. The FDE is defined as
\begin{equation}
    \text{FDE} = \frac{1}{N} \sum_{i=1}^{N} \sqrt{ (\hat{x}_{t+H}^{(i)} - x_{t+H}^{(i)})^2 + (\hat{y}_{t+H}^{(i)} - y_{t+H}^{(i)})^2 },
\end{equation}
where $N$ is the number of vehicles, $(\hat{x}_{t+H}^{(i)}, \hat{y}_{t+H}^{(i)})$ denotes the predicted coordinates of the $i$-th vehicle at the final time step, and $(x_{t+H}^{(i)}, y_{t+H}^{(i)})$ denotes the corresponding ground truth coordinates.

\subsubsection{\textbf{Vehicle Positioning via GNSS-RTK}}

To ensure precise synchronization between the physical environment and AdaPTwin, we assume that all vehicles are equipped with GNSS receivers supported by Real-Time Kinematic (RTK) positioning. RTK positioning utilizes carrier-phase measurements and correction streams from fixed base stations to achieve centimeter-level accuracy (typically 1–2 cm). This level of spatial fidelity is a prerequisite for site-specific ray-tracing. We consider a realistic end-to-end latency of $T_{P}$ = 100 ms for the position acquisition loop.

\subsubsection{\textbf{Continual Learning-Based Vehicle Trajectory Prediction (Online)}}

The continual learning-based vehicle trajectory prediction ensures that the offline-trained AI model can be adapted online to new environments and traffic patterns.

\textbf{Adapter-Based Tuning:}
Instead of retraining the entire Transformer backbone, which is computationally prohibitive on edge servers, we insert bottleneck Adapter modules after the feed-forward sub-layers of the Transformer blocks. An adapter consists of a down-projection $\mathbf{W}_{down} \in \mathbb{R}^{d_{model} \times d_{neck}}$, a non-linear activation $\phi(\cdot)$, and an up-projection $\mathbf{W}_{up} \in \mathbb{R}^{d_{neck} \times d_{model}}$, where $d_{neck} \ll d_{model}$. The output of an adapted layer is
\begin{equation}
    \mathbf{h}_{out} = \mathbf{h}_{in} + \mathbf{W}_{up} \phi(\mathbf{W}_{down} \mathbf{h}_{in}).
\end{equation}
During the online phase, the parameters of the pre-trained Transformer backbone are frozen. Only the Adapter parameters and the final regression head are updated. This ``Freeze-and-Adapt'' strategy significantly reduces the number of trainable parameters and the computational cost of gradient updates. To bound the drift of the adapted parameters from the offline solution, they are blended with their offline values after every update, $\boldsymbol{\theta} \leftarrow (1-\beta)\boldsymbol{\theta}_{\mathrm{ft}} + \beta\boldsymbol{\theta}_{\mathrm{off}}$, with $\beta = 0.5$.

\textbf{Experience Replay and Meta-Scheduling:}
To prevent catastrophic forgetting—where the model overfits to the most recent observations and forgets general motion priors—we employ an experience replay mechanism. A First-In-First-Out (FIFO) replay buffer stores a diversity of recent historical windows. At regular fine-tuning intervals (defined by parameter $\tau_{tune}$), the model is updated using a batch composed of both the latest observations and samples drawn from the replay buffer.

Furthermore, we utilize a meta-learning rate scheduler that exponentially decays the adaptation learning rate over time. This allows the model to adapt rapidly to the new environment initially and then stabilize its parameters as confidence in the site-specific distribution increases. The objective of online optimization aims to minimize the displacement error on the buffered trajectories, ensuring AdaPTwin maintains high prediction fidelity for real-time ray-tracing.

\vspace{-0.3cm}

\subsection{Network Configuration and Ray-Tracing}

Network configuration, ray-tracing, and CIR computation are executed via NVIDIA Sionna v2.0.1. To this end, a 3D virtual environment is constructed by (a) importing models of buildings, roads, and vehicles, (b) placing the RSU at specified locations, (c) assigning a receiver antenna to each vehicle, and (d) configuring the antennas of both transmitters and receivers based on parameters such as the number of elements, radiation pattern, polarization, and orientation. To synchronize the temporal and spatial domains, coordinate systems between the SUMO-generated trajectories and the 3D map are aligned before ingestion.

During online operation, the spatial coordinates and headings of all 3D vehicle meshes are continuously updated based on the trajectory prediction outputs to accurately reflect the upcoming physical network state. Following this geometry update, the hardware-accelerated solver utilizes the Shooting and Bouncing Rays (SBR) method to compute multipath interactions—including specular reflections, diffuse scattering, and diffraction. For every receiver (vehicle), the solver aggregates the contributions of all multipath components. The complex channel impulse response (CIR) is given by
    \begin{equation}
        h(\tau) = \sum_{k=1}^{N_{paths}} a_k e^{j \phi_k} \delta(\tau - \tau_k),
    \end{equation}
    where $a_k$, $\phi_k$, and $\tau_k$ represent the amplitude, phase, and delay of the $k$-th propagation path, respectively.

\vspace{-0.3cm}

\section{Development and Integration of AdaPTwin Algorithms}

\subsection{Phase 1: Cloud-Based Fidelity Policy Optimization}

Phase 1 runs in the cloud every $T_{\text{update}}$ (e.g., 10\,min), which suffices because macroscopic trends, such as the average vehicle density, change gradually. Its primary objective is to derive an optimal \textit{Fidelity Policy Map} $\pi^*$ that maps the current network context to an NDT fidelity configuration $C^*$, ensuring that the edge can complete the entire NDT-enabled proactive RRM loop within one slot. Here, the network context is defined as the vehicles' trajectory history collected over the preceding macroscopic time interval $T_{\text{update}}$ (e.g., 10\,min), consisting of time-indexed vehicle IDs and spatial coordinates. The context is thus accumulated over the window rather than a one-shot snapshot. From this history, the cloud extracts active vehicles and their positions at the timestamp with the largest number of vehicles, which makes the latency estimate conservative, derives vehicle orientations from sequential coordinates, and evaluates the scene update overhead $T_O$ across consecutive steps, enabling accurate fidelity evaluation against the high-fidelity reference NDT. Note that the cloud--edge exchange (a trajectory log and a six-element configuration vector once per $T_{\text{update}}$) is not on the per-slot critical path, and faster traffic changes can be handled by event-triggered recalibration.

\subsubsection{\textbf{Fidelity Space}}
The channel model utilizes Shooting-and-Bouncing Rays (SBR). The fidelity of the simulation is governed by the configuration vector $C \in \mathcal{C}$, defined as
\begin{equation}
C = \langle D_{max}, N_{rays}, N_{paths}, DR, D, F_{V} \rangle.
\end{equation}
Its components are:
\begin{enumerate}
    \item \textit{Ray Tracing Max Depth  ($D_{max}$):} Determines the maximum number of interactions (reflections/scattering) a ray can undergo before termination.
    \item \textit{No. of Sample Rays per Source in Ray Tracing ($N_{rays}$):} The angular resolution of ray launching, critical for finding paths in complex clutter.
    \item \textit{Max No. of Paths per Source in Ray Tracing ($N_{paths}$):} The maximum number of paths retained per source for channel reconstruction.
    \item \textit{Diffuse Reflection ($DR$):} A boolean flag enabling scattering from rough surfaces based on the Lambertian model.
    \item \textit{Diffraction ($D$):} A boolean flag enabling diffraction, i.e., the bending of waves around obstacle edges and through apertures. As it requires edge detection and substantially increases the computational cost, it is enabled only for cloud-based high-fidelity reference NDT and benchmarking.
    \item \textit{Fidelity of 3D Vehicles ($F_{V}$):} Fidelity of simplified 3D models of vehicles.
\end{enumerate}

In our experiments, $\mathcal{C}_{edge}$ contains $420$ configurations: $D_{max} \in \{1, \dots, 10\}$, $21$ values of $N_{rays} \in [10^2, 10^6]$ with $N_{paths} = 10 N_{rays}$, diffuse reflection on, diffraction off, and two low-poly mesh levels $F_V \in \{30, 50\}$, the two coarsest of the four simplified mesh levels (a larger $F_V$ denotes a coarser mesh).

\subsubsection{\textbf{High-Fidelity Reference and Latency Benchmarking}}
The cloud generates the high-fidelity reference data using AdaPTwin with the maximum realism settings available to our solver ($D_{max} = 10$, $N_{rays} = 10^6$, $N_{paths} = 10^7$, the most detailed vehicle meshes, and both diffuse reflection and diffraction enabled). We call it the \textit{high-fidelity reference NDT} rather than ground truth, since it is the most accurate channel our solver can synthesize.
The cloud also executes the \textbf{RRM optimization solver} (described in Phase 2) using the reference channels. This execution is not for scheduling, but to benchmark the computational cost of the RRM algorithm ($T_{RRM}^{C}$), which is a significant component of the total latency budget.

\subsubsection{\textbf{Fidelity Selection}}
To determine which configurations are feasible at the edge, the edge execution times are estimated from cloud measurements through $\kappa$, and a latency budget is derived from the prediction horizon $T_{H}$.

The effective time budget available for ray-tracing, denoted as $T_{Budget}$, is derived by subtracting the mandatory non-ray-tracing latencies from the horizon
\begin{equation}
    T_{Budget} = T_{H} - T_{P} - T_{TP} - \kappa (T_{RRM}^{C} + T_{O}) - T_{G}, \label{eq:budget}
\end{equation}
where $T_{P}$ is the end-to-end
latency for the position acquisition via GNSS-RTK, $T_{TP}$ is the end-to-end latency of trajectory prediction, including the online continual-learning update, measured offline on the edge platform, $T_{O}$ accounts for scene graph updates, and $T_{G}$ is a guard time that absorbs latency fluctuations caused by changes in the vehicle density. Since $T_P$ and $T_{TP}$ are edge-side quantities, only the cloud-measured $T_{RRM}^{C}$ and $T_O$ are scaled by $\kappa$.

As illustrated in Algorithm \ref{alg:fidelity}, the cloud then evaluates a search space of fidelity configurations $\mathcal{C}_{edge}$. For each candidate $C_k$, the estimated edge ray-tracing latency is $\tau_{est}(C_k) = \kappa \cdot \tau_{cloud}(C_k)$. The optimal configuration $C^*$ is selected to minimize the path gain RMSE with respect to the high-fidelity reference NDT, subject to the feasibility constraint
\begin{align}
    C^* = \arg \min_{C_k \in \mathcal{C}_{edge}} \quad & \text{RMSE}(\text{CIR}(C_k), \text{CIR}(C_{GT})) \\
    \text{s.t.} \quad & \tau_{est}(C_k) \leq T_{Budget}. \label{eq:feas}
\end{align}

\begin{algorithm}[!t]
\caption{Adaptive Fidelity Selection Policy}
\label{alg:fidelity}
\begin{algorithmic}[1]
\Require High-Fidelity Reference Config $C_{GT}$, Candidate Set $\mathcal{C}_{edge}$, Horizon $T_{H}$, Latency Ratio $\kappa$
\Ensure Optimal Edge Configuration $C^*$

\Statex \textbf{1. Benchmarking}
\State Generate Reference CIR $\mathbf{H}_{GT}$ using $C_{GT}$
\State Measure RRM time on Cloud: $T_{RRM}^{C} \gets \text{Time}(\text{RRM}(\mathbf{H}_{GT}))$
\State Measure scene-update overhead $T_{O}$ over consecutive steps \State Calculate Edge Time Budget $T_{Budget}$ via \eqref{eq:budget}:
\State $T_{Budget} \gets T_{H} - T_{P} - T_{TP} - \kappa(T_{RRM}^{C} + T_{O}) - T_{G}$

\Statex \textbf{2. Feasibility Search}
\State $\mathcal{E}_{min} \gets \infty$, $C^* \gets \emptyset$
\For{each candidate $C_k \in \mathcal{C}_{edge}$}
    \State Measure Cloud ray-tracing latency $\tau_{cloud}(C_k)$
    \State Estimate Edge latency: $\tau_{est}(C_k) \gets \kappa \cdot \tau_{cloud}(C_k)$
    
    \If{$\tau_{est}(C_k) \leq T_{Budget}$} \Comment{Feasibility Constraint \eqref{eq:feas}}
        \State Compute CIR $\mathbf{H}_k$ using configuration $C_k$
        \State Calculate Error $\mathcal{E}_k \gets \text{RMSE}(\mathbf{H}_k, \mathbf{H}_{GT})$
        \If{$\mathcal{E}_k < \mathcal{E}_{min}$}
            \State $\mathcal{E}_{min} \gets \mathcal{E}_k$
            \State $C^* \gets C_k$
        \EndIf
    \EndIf
\EndFor
\Statex \textbf{3. Graceful Degradation} \If{$C^* = \emptyset$} \Comment{No feasible configuration}    
\State $C^* \gets \arg\min_{C_k \in \mathcal{C}_{edge}} \tau_{est}(C_k)$ \Comment{Leanest config.}  \EndIf \State \Return $C^*$
\end{algorithmic}
\end{algorithm}

\vspace{-0.3cm}

\subsection{Phase 2: Edge-Based Proactive RRM}
This phase executes in every slot on the edge server co-located with the RSU. It performs steps (5)--(8) of Section IV-A with the fidelity $C^*$ selected in Phase 1, and the predicted CIRs are used by the RRM algorithm described below for vehicle--RSU association and beam selection.

\subsubsection{\textbf{Interference Characterization and Optimal Baseline}}

Let $\mathbf{w} = [w_1, w_2, \dots, w_B] \in \mathcal{W}^B$ denote a global network beam configuration, where $w_b \in \mathcal{W}$ is the active beam index at RSU $b$. To efficiently evaluate the exact SINR established in (1) during real-time edge execution, we isolate the predicted beamforming gain derived from the instantaneous CIR provided by AdaPTwin, defined mathematically as $G_{u,b,k} = |\mathbf{h}_{u,b}^H \mathbf{f}_k|^2$. Consequently, the aggregate inter-cell interference experienced by user $u$ can be dynamically calculated as $\sum_{j \in \mathcal{B} \setminus \{b\}} P_{tx} G_{u,j,w_j}$, which captures the exact electromagnetic footprint of all neighboring RSUs transmitting on their specific active beams.

An exhaustive search over all $W^B$ beam states, where $W$ is the codebook size and $B$ the number of RSUs, attains the optimum but exceeds the edge latency budget; it is therefore used only as an upper-bound benchmark for the heuristic proposed below.

\subsubsection{\textbf{Real-Time Heuristic RRM Algorithm}}

To satisfy the strict latency constraint defined in \eqref{eq:c5}, we propose a low-complexity Iterative Coordinate Descent (ICD) algorithm augmented with a random multi-start framework to escape local optima.

The proposed algorithm compares candidate solutions by the pair $\text{Obj} = (f_{\text{eval}}, \gamma_{\text{worst}})$ of the score \eqref{eq:score}, which penalizes users below $\gamma_{\min}$, and the worst-case SINR of the associated users: $\text{Obj}_a \succ \text{Obj}_b$ if $f_{\text{eval},a} > f_{\text{eval},b}$, or if $f_{\text{eval},a} = f_{\text{eval},b}$ and $\gamma_{\text{worst},a} > \gamma_{\text{worst},b}$. Hence, a candidate is accepted if it improves the score, and the worst-case SINR only resolves exact ties. With the fidelity configuration $C^*$ determined by the cloud, the penalty term $-\lambda \mathcal{E}(C^*)$ in \eqref{eq:obj} acts as an additive constant, simplifying edge optimization to proportional fairness maximization. In Algorithm~\ref{alg:rrm}, the subroutine $\text{Evaluate}(\mathbf{w})$ maps candidate beam state $\mathbf{w}$ to user associations $\mathbf{X}$ by associating users to the active beam offering the highest predicted SINR up to per-RSU load limit $L$, adhering to constraints \eqref{eq:c1}--\eqref{eq:load}. Users below $\gamma_{\min}$ are in outage \eqref{eq:c_sinr}, and $\text{Evaluate}(\mathbf{w})$ returns the score \eqref{eq:score} of this association, with the worst-case user SINR resolving ties. As illustrated in Algorithm \ref{alg:rrm}, this algorithm proceeds as follows:

\begin{enumerate}
\item \textbf{Multi-Start Initialization:} 
The algorithm runs for a predefined number of restarts. In the first restart, it initializes the beam configuration $\mathbf{w}^{(0)}$ by selecting the beam with the maximum single-user channel gain for each RSU, ignoring inter-cell interference. In subsequent restarts, the initial beam configuration is completely randomized.

\item  \textbf{Iterative Coordinate Descent:}
In each iteration, the RSUs are updated sequentially in a random permutation order to prevent bias. For a specific RSU $b$, the beams of all other RSUs $\mathbf{w}_{-b}$ are held fixed. The algorithm evaluates all beams $k \in \mathcal{W}$ for RSU $b$ and selects the beam $w_b^*$ that yields the highest objective value given the fixed interference pattern from neighbors.

\item  \textbf{Exact Interference Evaluation:}
The algorithm computes the exact SINR and the soft satisfaction score \eqref{eq:score} for every candidate update. If two beams yield the same aggregate satisfaction score, the worst-case SINR across all associated users is used as a tie-breaker. The descent loop continues until no improvements can be found in a full pass, and the best configuration across all restarts is returned.
\end{enumerate}

\textit{Soft satisfaction score.} Algorithm~\ref{alg:rrm} maximizes $f_{\text{eval}}(\mathbf{w}) = \sum_{u} s_u(\mathbf{w})$ over the associated users, with
\begin{equation}
s_u(\mathbf{w}) = \begin{cases} \ln\!\big(\max\{\log_2(1+\gamma_{u,b,\mathbf{w}}),\epsilon\}\big) + c_0, & \gamma_{u,b,\mathbf{w}} \geq \gamma_{\min},\\ -M, & \text{otherwise}, \end{cases} \label{eq:score}
\end{equation}
where $\epsilon = 10^{-9}$, $c_0 = 2$, and $M = 100$. Equivalently, \[ f_{\text{eval}}(\mathbf{w}) = \sum_{u \in \mathcal{S}(\mathbf{w})} \big(\ln(\log_2(1+\gamma_{u,b,\mathbf{w}})) + c_0\big) - M\,|\mathcal{O}(\mathbf{w})|, \] where $\mathcal{S}(\mathbf{w})$ and $\mathcal{O}(\mathbf{w})$ are the associated users that meet and miss $\gamma_{\min}$, respectively: the proportional-fairness term of \eqref{eq:obj} over the users meeting $\gamma_{\min}$, shifted by $c_0$ per user, minus a penalty $M$ for every user in outage. Both terms are computed together in every call of $\text{Evaluate}(\mathbf{w})$, so they are not optimized one after the other. However, bringing one more user above $\gamma_{\min}$ raises the score by at least $M + c_0 + \ln(\log_2(1+\gamma_{\min})) \approx 100.9$, whereas the proportional-fairness term of a served user varies by only about $4$ over the admissible SINRs. Algorithm~\ref{alg:rrm} therefore primarily reduces the number of users in outage and, among beam states with the same number of users in outage, maximizes the proportional fairness. The offset $c_0$ also prevents \eqref{eq:obj} from dropping admissible users with a spectral efficiency below $1$\,bps/Hz, whose logarithm is negative.  If a configuration results in an overloaded RSU (serving $>L$ users), the user association step employs a sub-greedy approach where only the top $L$ users with the highest spectral efficiency are served, and the remaining users are dropped for that specific beam configuration. This ensures the load constraint \eqref{eq:load} is strictly satisfied during the optimization.

\begin{algorithm}[!t]
\caption{Heuristic RRM via Multi-Start Iterative Coordinate Descent}
\label{alg:rrm}
\begin{algorithmic}[1]
\Require Predicted Channels $\mathbf{H}$, Codebook $\mathcal{W}$, Load Limit $L$, Num Restarts $N_{res}$, Max Iters $N_{iter}$
\Ensure Best Global Beam Vector $\mathbf{w}^*$, User Association $\mathbf{X}^*$

\State $\text{Obj}^* \gets (-\infty, -\infty)$ \Comment{$\text{Obj} = (f_{\text{eval}}, \gamma_{\text{worst}})$}
\For{$r = 0$ to $N_{res} - 1$}
    \Statex \quad \textbf{1. Initialization}
    \If{$r = 0$}
        \State $w_b \gets \arg\max_{k \in \mathcal{W}} \max_u | \mathbf{h}_{u,b}^H \mathbf{f}_k |^2, \forall b \in \mathcal{B}$
    \Else
        \State $w_b \gets \text{Random}(\mathcal{W}), \forall b \in \mathcal{B}$
    \EndIf
    \State $\text{Obj}_{curr} \gets \text{Evaluate}(\mathbf{w})$
    \Statex \quad \textbf{2. Iterative Coordinate Descent}
    \For{$t = 1$ to $N_{iter}$}
        \State $improved \gets \text{False}$
        \For{each RSU $b \in \text{RandomPermutation}(\mathcal{B})$}
            \State $\mathbf{w}_{-b} \gets \{w_j\}_{j \neq b}$
            \State $k_{best} \gets w_b$, $\text{Obj}_{local} \gets \text{Obj}_{curr}$
            \For{each candidate $k \in \mathcal{W} \setminus \{w_b\}$}
                \State $\text{Obj}_{cand} \gets \text{Evaluate}(k \cup \mathbf{w}_{-b})$
                \If{$\text{Obj}_{cand} \succ \text{Obj}_{local}$}
                    \State $\text{Obj}_{local} \gets \text{Obj}_{cand}$
                    \State $k_{best} \gets k$
                    \State $improved \gets \text{True}$
                \EndIf
            \EndFor
            \State $w_b \gets k_{best}$, $\text{Obj}_{curr} \gets \text{Obj}_{local}$
        \EndFor
        \If{\textbf{not} $improved$} 
            \State \textbf{break} 
        \EndIf
    \EndFor
    \Statex \quad \textbf{3. Global Update}
    \If{$\text{Obj}_{curr} \succ \text{Obj}^*$}
        \State $\text{Obj}^* \gets \text{Obj}_{curr}$
        \State $\mathbf{w}^* \gets \mathbf{w}$
    \EndIf
\EndFor
\Statex \textbf{4. Final User Association (Sub-Greedy)}
\For{each RSU $b \in \mathcal{B}$ with active beam $w_b^*$}
    \State Calculate SINR $\gamma_{u,b}$ for all users $u$
    \State Sort users by spectral efficiency (SE)
    \State $x_{u,b} \gets 1$ for top-$L$ users with highest SE, $0$ for others
\EndFor
\State \Return $\mathbf{w}^*, \mathbf{X}^*$
\end{algorithmic}
\end{algorithm}

\subsubsection{\textbf{Complexity and Scalability Analysis}}

Whereas exhaustive search evaluates $\mathcal{O}(W^B)$ beam states, each descent pass of the proposed algorithm evaluates $BW$ beam states, each at a cost of $\mathcal{O}(UB\log(UB))$ for the SINR computation and the greedy association. Its worst-case complexity is thus $\mathcal{O}(N_{res} N_{iter} B W \cdot UB\log(UB))$, i.e., polynomial rather than exponential in $B$.

\vspace{-0.3cm}
\section{Performance Evaluation}
In this section, we evaluate AdaPTwin against the high-fidelity reference NDT, reactive single- and multi-fidelity NDTs, the stochastic 3GPP channel model, predictive single-fidelity NDTs, and a non-adaptive predictive multi-fidelity NDT in different cities and traffic densities. The evaluated metrics are the path gain RMSE, end-to-end latency, outage probability, sum rate, blockage prediction error rate, and FDE. Sections VII-B to VII-D compare AdaPTwin with reactive, single-fidelity, and non-adaptive multi-fidelity NDTs, Section VII-E presents ablation studies and a sensitivity analysis, Section VII-F evaluates the optimality and scalability of the RRM and of the two-timescale decomposition, and Section VII-G assesses the statistical reliability of the results.

\vspace{-0.3cm}

\subsection{Simulation Setup}
The proposed AdaPTwin framework is written in Python, and we used a Linux system with an NVIDIA RTX A6000 GPU to execute the code. As the same computation engine was used for both phases, $\kappa$ (edge-to-cloud Latency Ratio) was considered 1 in simulations. Setting $\kappa = 1$ corresponds to edge servers with GPU resources comparable to the cloud; for $\kappa > 1$, $T_{Budget}$ in \eqref{eq:budget} shrinks and Phase~1 selects a leaner configuration, trading a slightly higher RMSE for meeting the deadline (Section VII-E). Two complex urban areas with several intersections in Downtown Vancouver, Canada, and Downtown Ottawa, Canada, were selected for evaluating AdaPTwin (cf. Supplementary Material, which provides visualizations of their 3D models in Blender and corresponding SUMO road networks). These environments are challenging for vehicular networks, since dense high-rise geometry and frequent intersections produce rapid LOS transitions and rich multipath. To rigorously test the framework's adaptivity and scalability, scenarios with varying traffic densities (medium and high) were simulated. The average number of vehicles in the environment in each time slot for the different scenarios is as follows: 25 for medium density of Ottawa, 40 for high density of Ottawa, 35 for medium density of Vancouver, and 50 for high density of Vancouver. For the Vancouver scenarios, the Transformer is trained offline on the SUMO dataset of Ottawa, and for the Ottawa scenarios on that of Vancouver, i.e., it is always deployed on a road network and on trajectories not seen in training. It is not retrained for the evaluated scenario: only its adapters and regression head are updated online by the continual-learning mechanism of Section V-B.  

\textit{Baselines.} All baselines are re-implemented in the same pipeline (scenes, RSUs, antennas, solver, reference NDT, RRM optimizer, and deadline rule), so that they differ only in the mechanism under study. The reactive twins \cite{Demir2023Reactive, Salehi2024Multiverse} ray-trace the positions of the previous slot, the framework of Pegurri \textit{et al}. \cite{Pegurri2026Proactive} uses our Transformer instead of its Kalman filter, which favors it, and each fidelity level of the multi-fidelity baselines is reported separately. The 3GPP UMi model \cite{3GPPTR38901UMi} is included not as a competitive site-specific baseline but as the low-cost option for channel synthesis; evaluated at the accurately known vehicle positions of the previous slot, it shows how much of the sum rate a non-site-specific model loses. Static map errors are common to AdaPTwin and all NDT baselines, which share the same scene, so they favor none of them.

\textit{Deadline rule.} A predictive decision with latency $\tau > T_H$ occupies part of the next slot and keeps the fraction $\max\{0, 2 - \tau/T_H\}$ of its sum rate, whereas a reactive decision, computed within the slot it serves, keeps $\max\{0, 1 - \tau/T_H\}$. Outage is reported as measured.

We considered 3D models of different types of vehicles, including car, bus, and box truck, and chose the appropriate radio materials of the objects in the environment as follows: concrete for the buildings, medium-dry ground for roads and the ground, and metal for the vehicles. For adaptive fidelity selection, the maximum number of simulated paths per source ($N_{paths}$) was strictly coupled to the number of sample rays ($N_{rays}$), utilizing a quasi-logarithmic sweep from $10^2$ to $10^6$ for $N_{rays}$, with $N_{paths}$ set to 10 times $N_{rays}$. The ray tracing parameters for non-adaptive frameworks that use predefined levels of fidelity were selected as follows. The max depth was selected as 3, 6, and 10 for low, medium, and high fidelity, respectively. The number of sample rays per source was set to $10^{3}$, $10^{4}$, and $10^{6}$ for low, medium, and high fidelity, respectively, and the maximum number of paths per source was set to 10 times the number of sample rays per source. Low, medium, and high fidelity of low-poly 3D models of vehicles were considered for low, medium, and high fidelity, respectively.

The per-slot deadline (real-time deadline) is determined by the prediction horizon $T_H$, which is set to $1$\,s in the simulations. For evaluating outage probability, the target minimum SINR ($\gamma_{min}$) is bounded to -6\,dB \cite{3GPPTR36885SINR}. For each RSU, the codebook contains $W = 16$ beams spanning $\pm 60^\circ$ in azimuth, and the maximum load capacity per RSU ($L$) is constrained to 100 vehicles to simulate realistic limitations. The detailed parameters of the vehicle trajectory prediction engine and the parameters of channel prediction and RRM engines are summarized in Table~\ref{tab:traj_params} and Table~\ref{tab:radio_params}, respectively.

\setlength{\textfloatsep}{0pt} 
\begin{table}[!t]
\scriptsize
\centering
\caption{Parameters of the Vehicle Trajectory Prediction Engine}
\label{tab:traj_params}
\renewcommand{\arraystretch}{1.2}
\begin{tabular}{ll}
\toprule
\textbf{Parameter} & \textbf{Value} \\
\midrule
Prediction Horizon & 1 s \\
Transformer History Length & 3 s \\
LSTM History Length & 2 s \\
KF Tracking History & 30 Frames \\
Vancouver (High Density) No. of Vehicles & 17063 \\
Driving Time & 10 Hours \\
$d_{model}$ (Transformer) & 64 - 128 \\
No. of Heads (Transformer) & 4 \\
No. of Layers (Transformer) & 1 - 4 \\
$d_{ff}$ (Transformer) & 64 - 1024 \\
Dropout Rate (Transformer) & 0.02 - 0.25 \\
L2 Regularization (Transformer) & $1\times10^{-7}$ - $2\times10^{-6}$ \\
Batch Size (Transformer) & 32 - 1024 \\
Max. Learning Rate (Transformer) & 0.0001 \\
Min. Learning Rate (Transformer) & $2.8\times10^{-6}$ \\
No. of LSTM Units & 64 \\
LSTM Batch Size & 32 \\
LSTM Learning Rate & 0.001 \\
KF Process Noise Variance ($q_{var}$) & 1.0 \\
KF Measurement Noise Variance ($R$) & 0.001 \\
KF Initial Velocity Covariance ($P_{0(v_x, v_y)}$) & 1000.0 \\
No. of Epochs & 50 \\
Early Stopping Patience & 2 Epochs \\
Fine-Tuning Interval & 1 s \\
No. of Epochs for Fine-Tuning & 1 \\
\bottomrule
\end{tabular}
\end{table}
\vspace{-0.2cm}
\begin{table}[!t]
\scriptsize
\centering
\caption{Parameters of Channel Prediction and RRM Engines}
\label{tab:radio_params}
\renewcommand{\arraystretch}{1.2}
\begin{tabular}{ll}
\toprule
\textbf{Parameter} & \textbf{Value} \\
\midrule
Prediction Horizon & 1 s \\
Ray Tracing Max Depth & 1 -- 10 \\
No. of Sample Rays per Source & $10^2$ -- $10^6$ \\
Max No. of Paths per Source & $10^3$ -- $10^7$ \\
No. of Fidelity Levels of Simplified 3D Vehicles & $4$ \\
Diffuse Reflection & True \\
Carrier Frequency & 5.875 GHz \\
System Bandwidth & 20 MHz \\
Thermal Noise Density & -174 dBm/Hz \\
Receiver Noise Figure & 6.0 dB \\
No. of RSUs & 5 \\
No. of Vehicles (Test Phase) & 1380 \\
RSUs No. of Antennas & 16 \\
Vehicles No. of Antennas & 1 \\
RSU Antenna Pattern & Dipole \\
Vehicles Antenna Pattern & Isotropic \\
RSUs Transmitting Power & 44 dBm \\
RSUs Height & 10 m \\
RSUs Downtilt & $-11^\circ$ \\
Vancouver Area Size & $420\,\text{m}\times350\,\text{m}$ \\
Ottawa Area Size & $350\,\text{m}\times350\,\text{m}$ \\
RSUs Beam Codebook Size & 16 \\
Max. No. of Vehicles per RSU & 100 \\
Min. SINR ($\gamma_{min}$) & -6 dB \\
$\kappa$ (Edge-to-Cloud Latency Ratio) & 1 (swept over 1--1.5) \\
$T_{P}$ / $T_{TP}$ / $T_{G}$ & 100 / 120 / 50 ms \\
\bottomrule
\end{tabular}
\end{table}

\vspace{-0.3cm}

\subsection{Predictive NDT vs. Reactive NDT}
\begin{figure}[!t]
    \centering
    \subfloat[Latency]{
        \includegraphics[width=0.82\columnwidth]{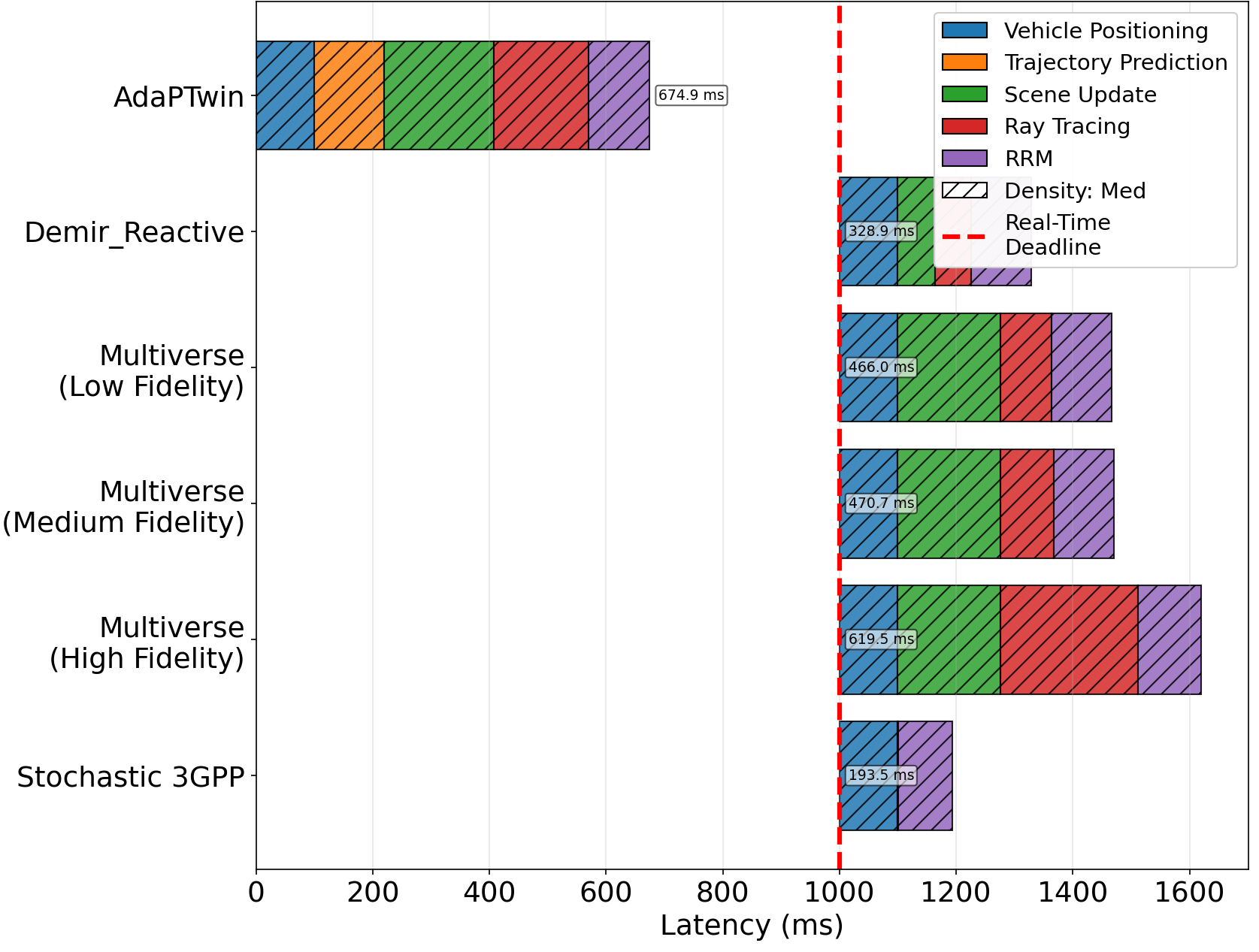}
        \label{fig:Predictive_Latency}
    }
    
    \subfloat[Outage Probability]{
        \includegraphics[width=1\columnwidth]{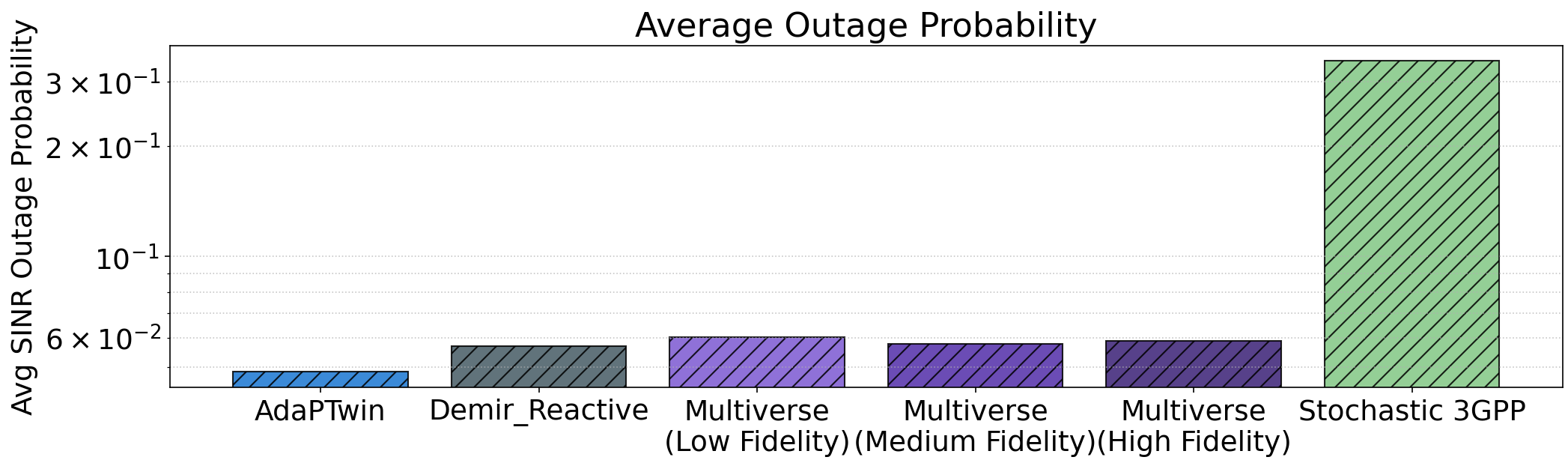}
        \label{fig:Predictive_Outage}
    }
    
    \subfloat[Sum Rate]{
        \includegraphics[width=0.95\columnwidth]{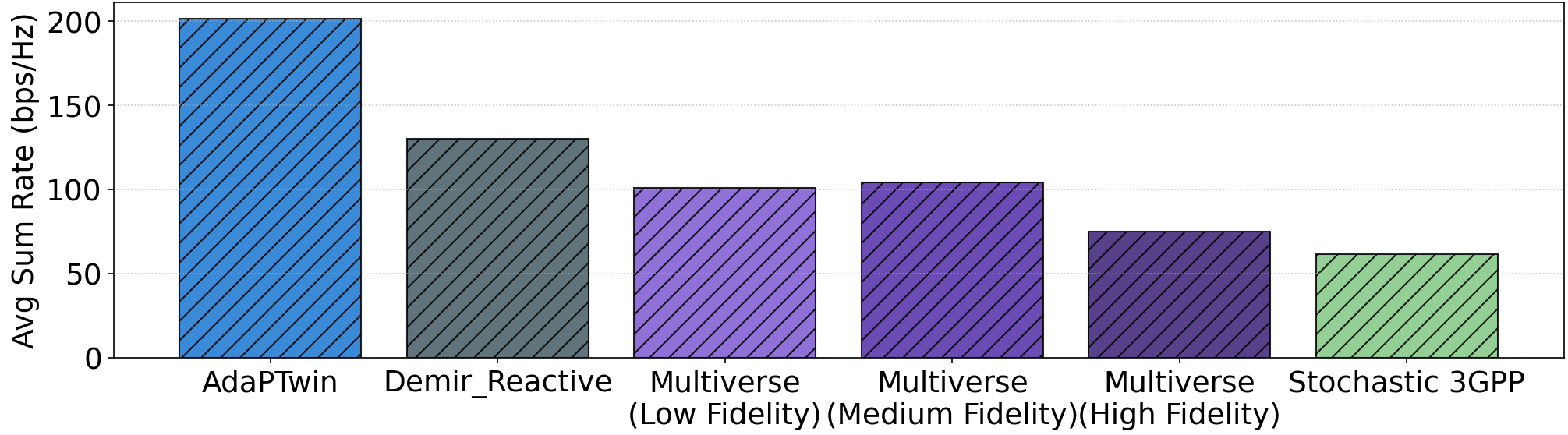}
        \label{fig:Predictive_Rate}
    }
    \caption{Comparison of predictive AdaPTwin framework against reactive single fidelity NDT, reactive multi-fidelity NDT (Multiverse), and the stochastic 3GPP channel model in terms of latency, outage probability, and sum rate.}
    \label{fig:Predictive}
    
\end{figure}

This subsection compares the predictive AdaPTwin framework against state-of-the-art reactive NDTs, namely the single-fidelity reactive NDT by Demir \textit{et al}. \cite{Demir2023Reactive}, the multi-fidelity reactive framework, called Multiverse \cite{Salehi2024Multiverse}, and the stochastic 3GPP UMi channel model \cite{3GPPTR38901UMi}. Figures $\,$\ref{fig:Predictive_Latency}, $\,$\ref{fig:Predictive_Outage}, and $\,$\ref{fig:Predictive_Rate} show this comparison in terms of latency, outage probability, and sum rate, respectively. Reactive NDTs ray-trace the current network state and overlook the latency of ray tracing and RRM; even Multiverse, which selects its fidelity based on available resources, cannot complete its decisions before the channel changes. Although it uses the accurately known vehicle positions of the previous slot, the stochastic 3GPP channel model cannot reproduce site-specific multipath, blockage, and angular structure; it therefore provides no user-specific beam-gain information and yields poor RRM decisions. The results indicate that, owing to its look-ahead strategy, AdaPTwin completes its decisions within the per-slot deadline in the evaluated scenarios, whereas the reactive NDTs do not. Furthermore, AdaPTwin achieves at least 55\% higher sum rate and 17\% lower outage probability compared to these reactive NDTs. 

\vspace{-0.3cm}
\begin{figure}[!t]
    \centering
    \subfloat[Latency]{
        \includegraphics[width=0.9\columnwidth]{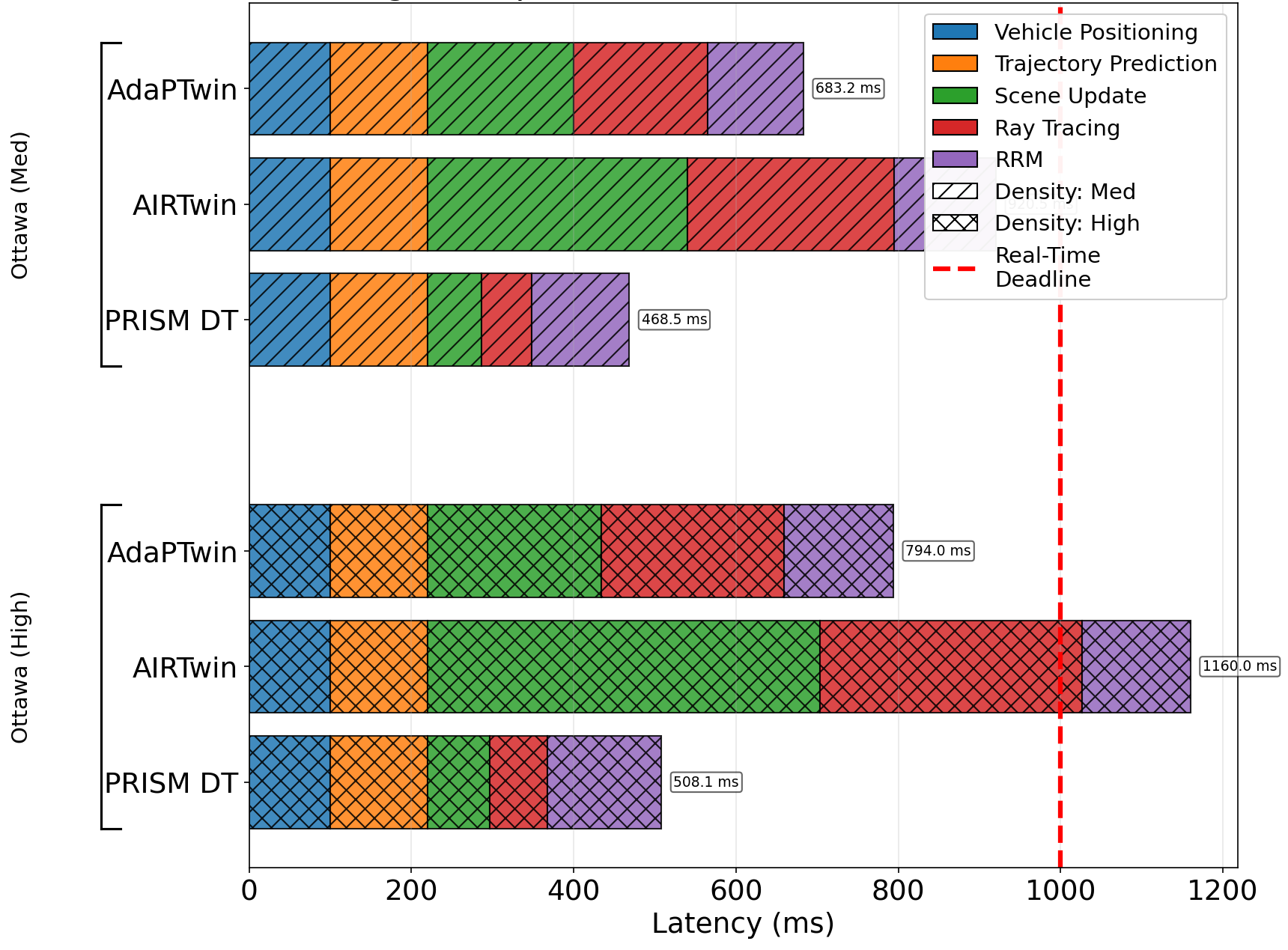}
        \label{fig:Multi_Latency}
    }
    
    \subfloat[Sum Rate]{
        \includegraphics[width=1\columnwidth]{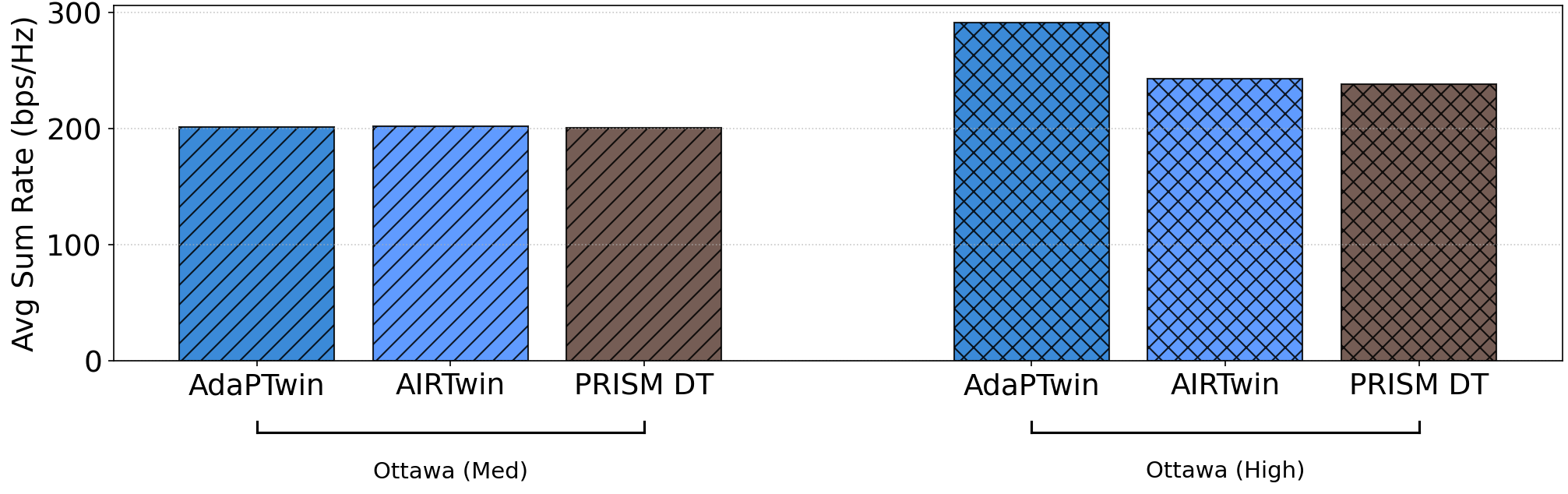}
        \label{fig:Multi_Rate}
    }
    \caption{Comparison of AdaPTwin with predictive single fidelity NDTs in terms of latency and sum rate.}
    \label{fig:Multi}
\end{figure}

\subsection{Multi-Fidelity vs. Single-Fidelity Predictive NDT}
To highlight the necessity of adaptive multi-fidelity, AdaPTwin is compared with PRISM DT \cite{Elloumi2025PRISM} (a predictive low-fidelity NDT) and AIRTwin \cite{Makvandi2026AIRTwin} (a predictive high-fidelity NDT). Figures $\,$\ref{fig:Multi_Latency} and $\,$\ref{fig:Multi_Rate} show the comparison of these models in terms of latency and sum rate under medium and high traffic densities in Ottawa. PRISM DT meets the deadline, but its low-fidelity configuration yields a poor sum rate at high vehicle density. AIRTwin meets the deadline and achieves a good sum rate at medium density, but at high density it does not complete its decisions within the time budget, so part of the next slot, which is reserved for data transmission, is occupied by decision-making and the sum rate degrades. In contrast, AdaPTwin adapts its fidelity to the traffic density, meeting the deadline and achieving the highest sum rate in all evaluated scenarios.

\vspace{-0.4cm}
\subsection{Adaptive vs. Non-Adaptive Predictive Multi-Fidelity NDT}
\begin{figure}[!t]
    \centering
    \subfloat[Latency]{
        \includegraphics[width=0.82\columnwidth]{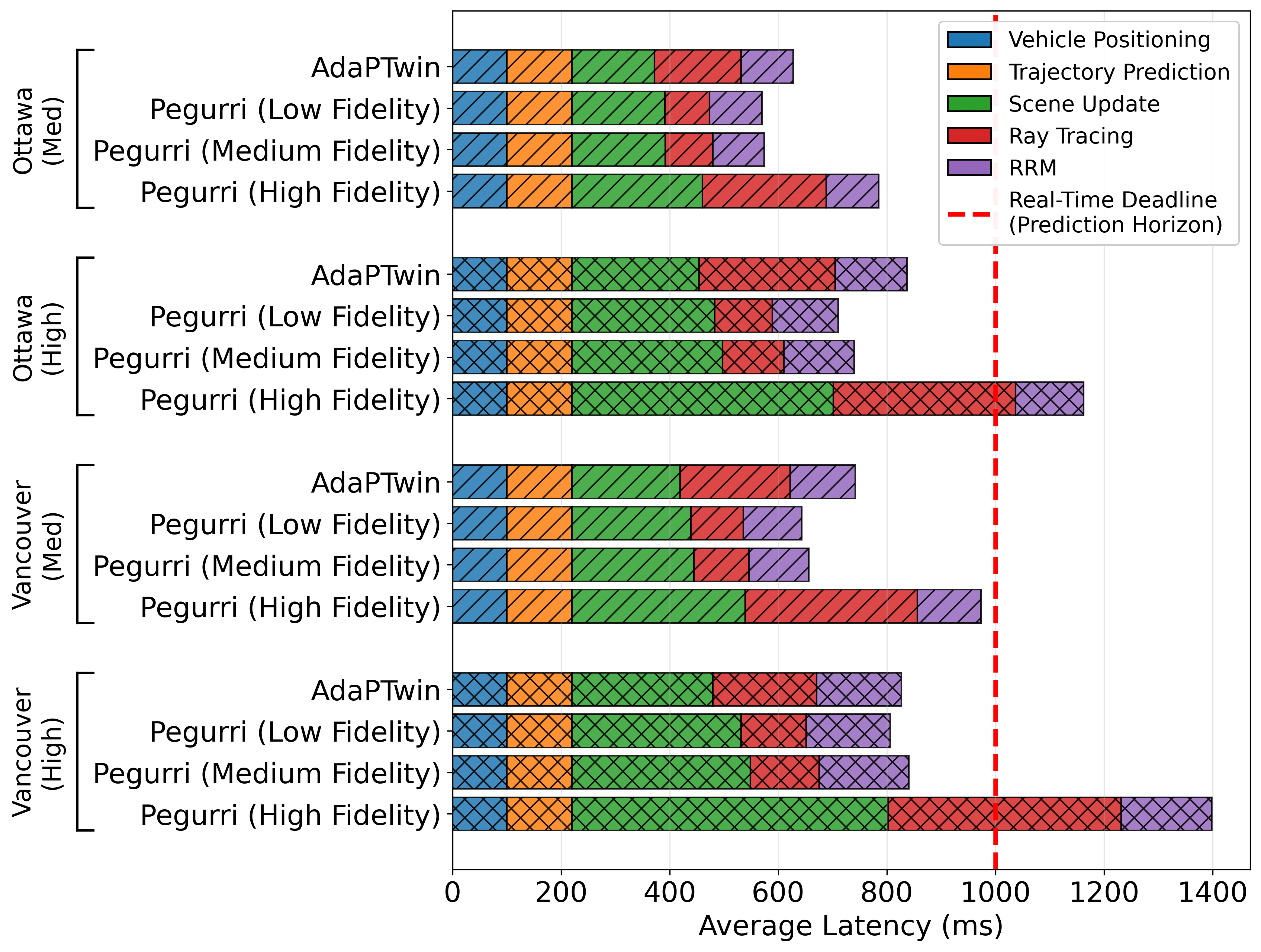}
        \label{fig:Adaptive_Latency}
    }
    
    \subfloat[RMSE]{
        \includegraphics[width=0.88\columnwidth]{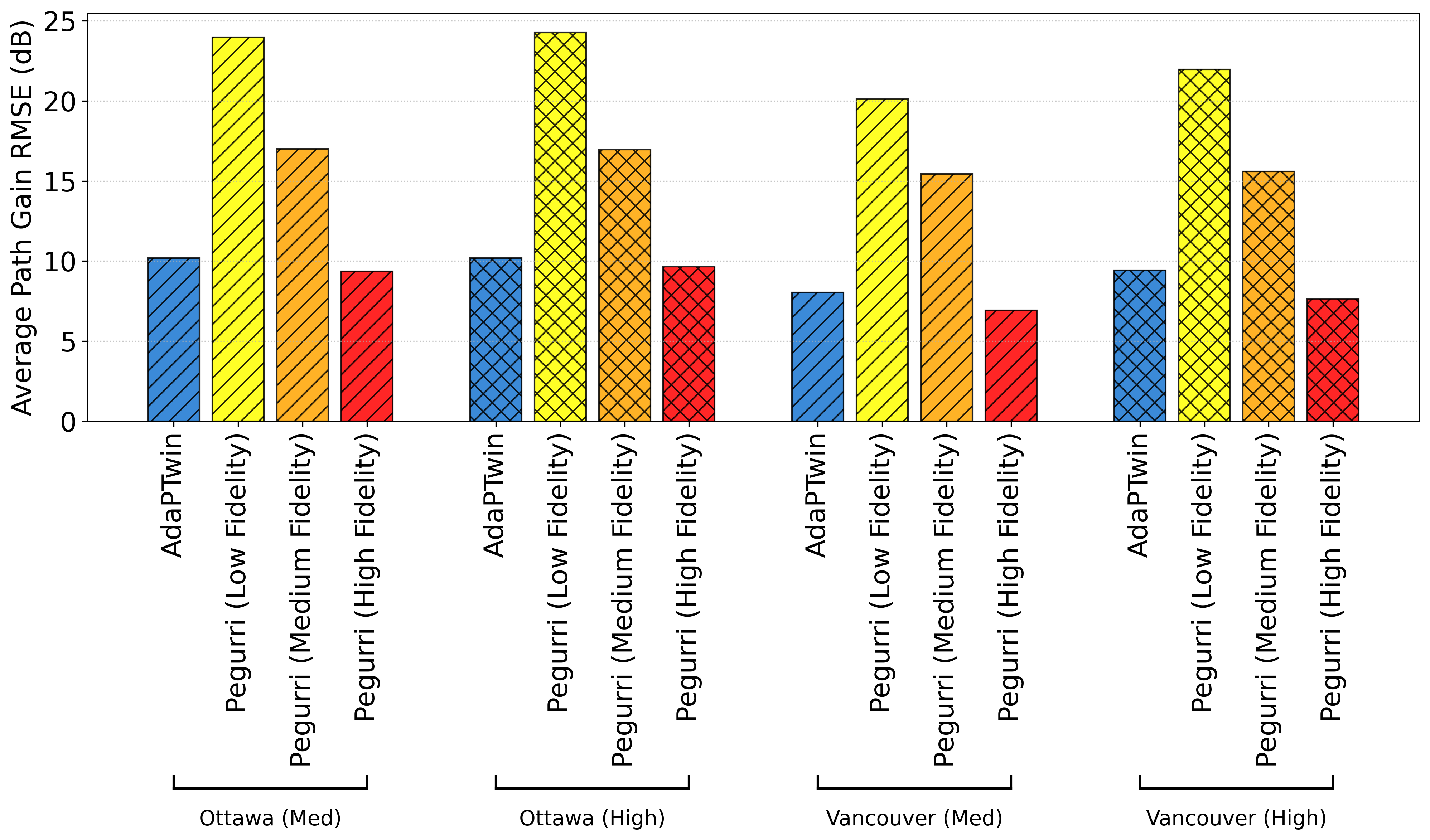}
        \label{fig:Adaptive_RMSE}
    }

        \subfloat[Outage Probability]{
        \includegraphics[width=0.88\columnwidth]{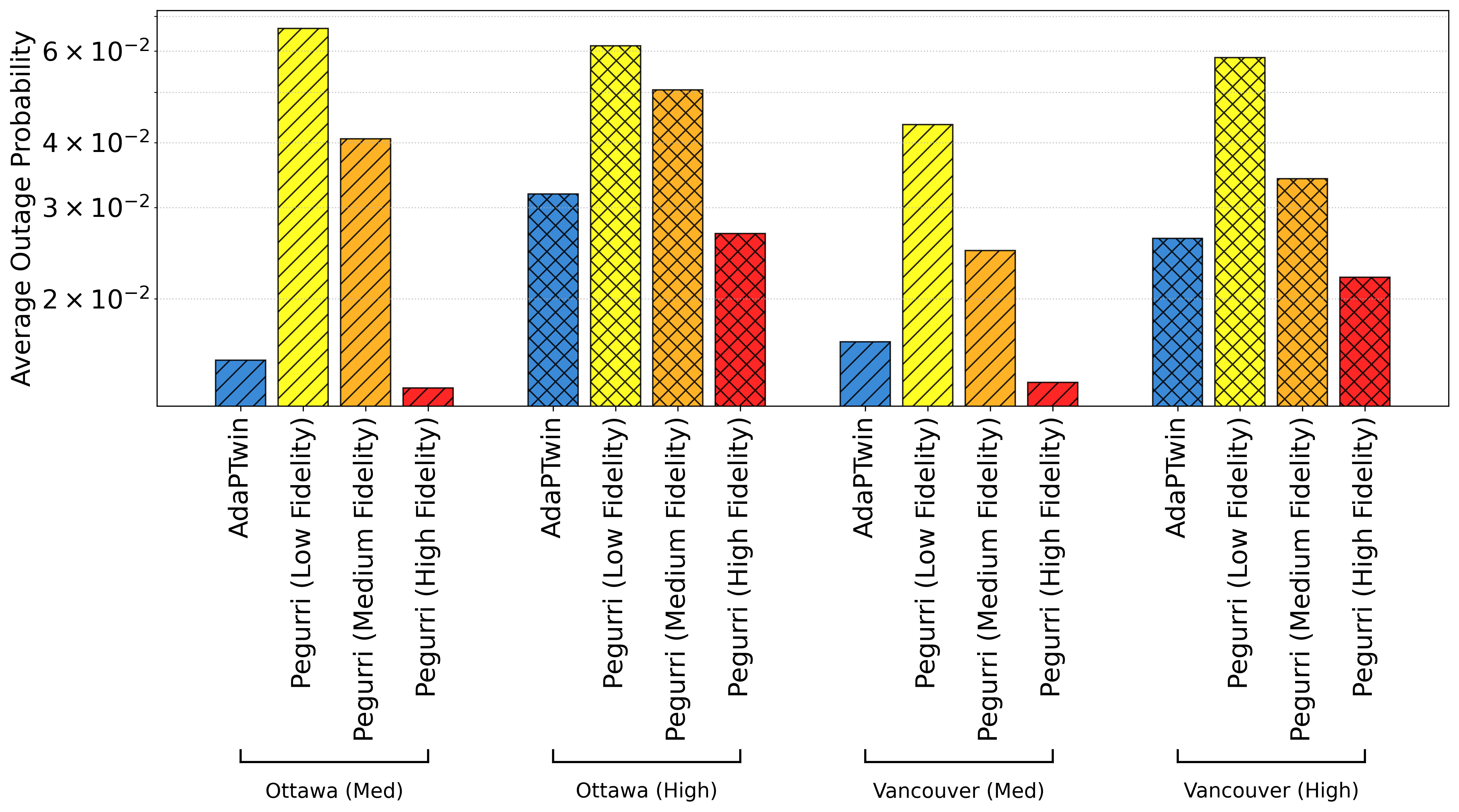}
        \label{fig:Adaptive_Outage}
    }

        \subfloat[Sum Rate]{
        \includegraphics[width=0.88\columnwidth]{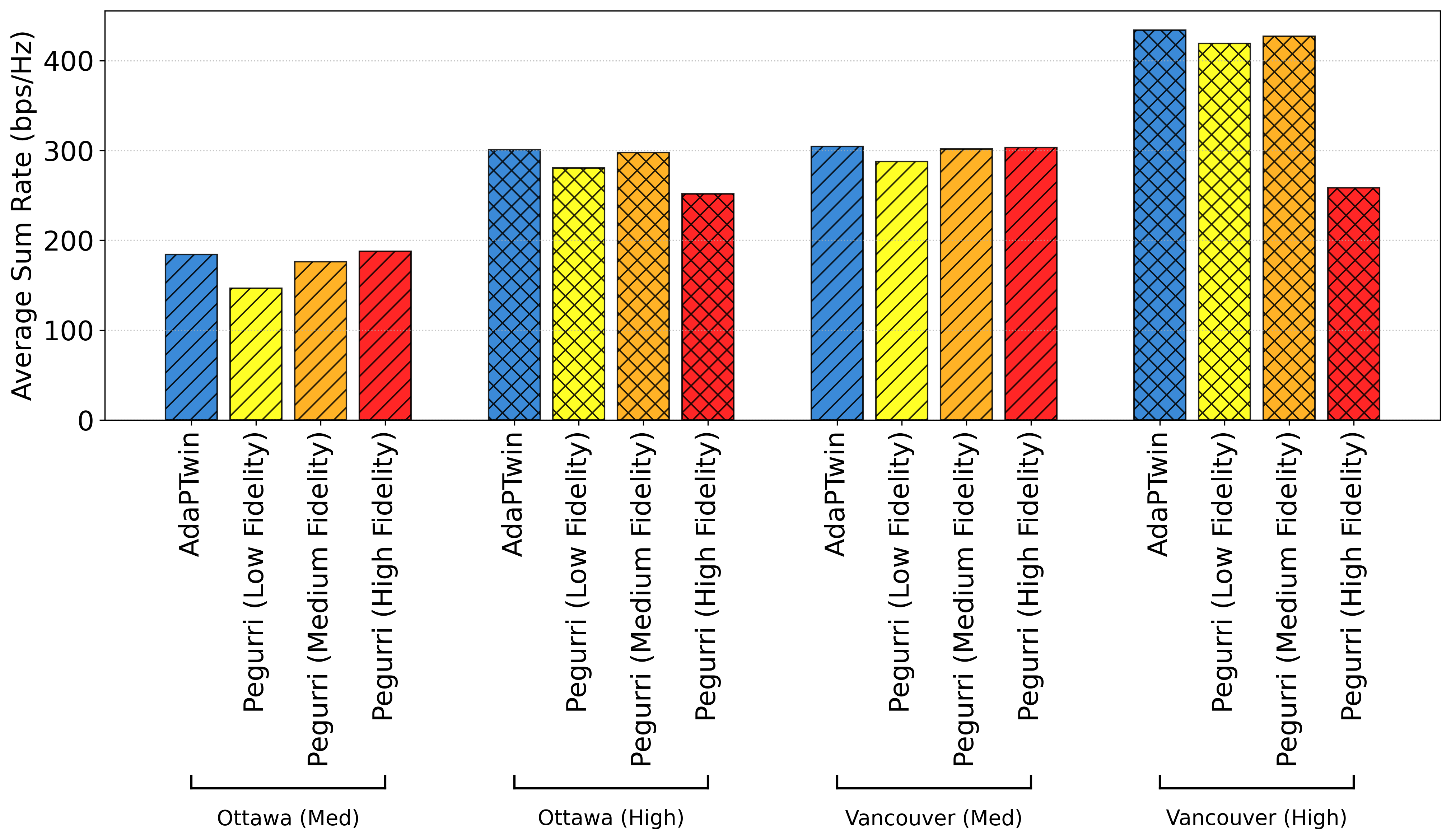}
        \label{fig:Adaptive_Rate}
    }
    \caption{Comparison of AdaPTwin with a non-adaptive predictive multi-fidelity NDT in terms of latency, path gain RMSE, outage probability, and sum rate.}
    \label{fig:Adaptive}
\end{figure}

We show the advantage of AdaPTwin against the non-adaptive predictive multi-fidelity framework by Pegurri \textit{et al}. \cite{Pegurri2026Proactive}. To isolate the fidelity policy, Pegurri's framework is equipped with our Transformer instead of its constant-velocity Kalman filter, which is less accurate (Section VII-E). Figures $\,$\ref{fig:Adaptive_Latency}, $\,$\ref{fig:Adaptive_RMSE}, $\,$\ref{fig:Adaptive_Outage}, and $\,$\ref{fig:Adaptive_Rate} show the comparison of these models in terms of latency, path gain RMSE, outage probability, and sum rate, respectively, under medium and high traffic densities in Vancouver and Ottawa.
\setlength{\textfloatsep}{0pt}

The results show that for the non-adaptive framework (Pegurri \textit{et al}.), the predefined levels of fidelity fail to adapt to different network settings. The low level of fidelity provides poor accuracy, outage probability, and sum rate, and the high level of fidelity fails to complete decision-making within the time budget, degrading the sum rate. Although medium fidelity outperforms low fidelity, it falls short of AdaPTwin, especially in terms of RMSE and outage probability. In contrast, AdaPTwin adapts to each scenario by selecting the most accurate latency-feasible configuration, achieving up to 70\% sum rate gain and 80\% outage probability reduction compared with the non-adaptive framework of Pegurri \textit{et al}., while its average latency remains below the per-slot deadline in every evaluated scenario.

\vspace{-0.4cm}
\subsection{Ablation Studies and Sensitivity Analysis}
\label{sec:ablation}
This subsection isolates the contribution of the main components of AdaPTwin (ablation) and relaxes its key assumptions (sensitivity analysis).

\subsubsection{\textbf{3D Modeling of Vehicles and Dynamic Blockage}}
\begin{figure}[!t]
    \centering
    \subfloat[Blockage Prediction Error Rate]{
        \includegraphics[width=1\columnwidth]{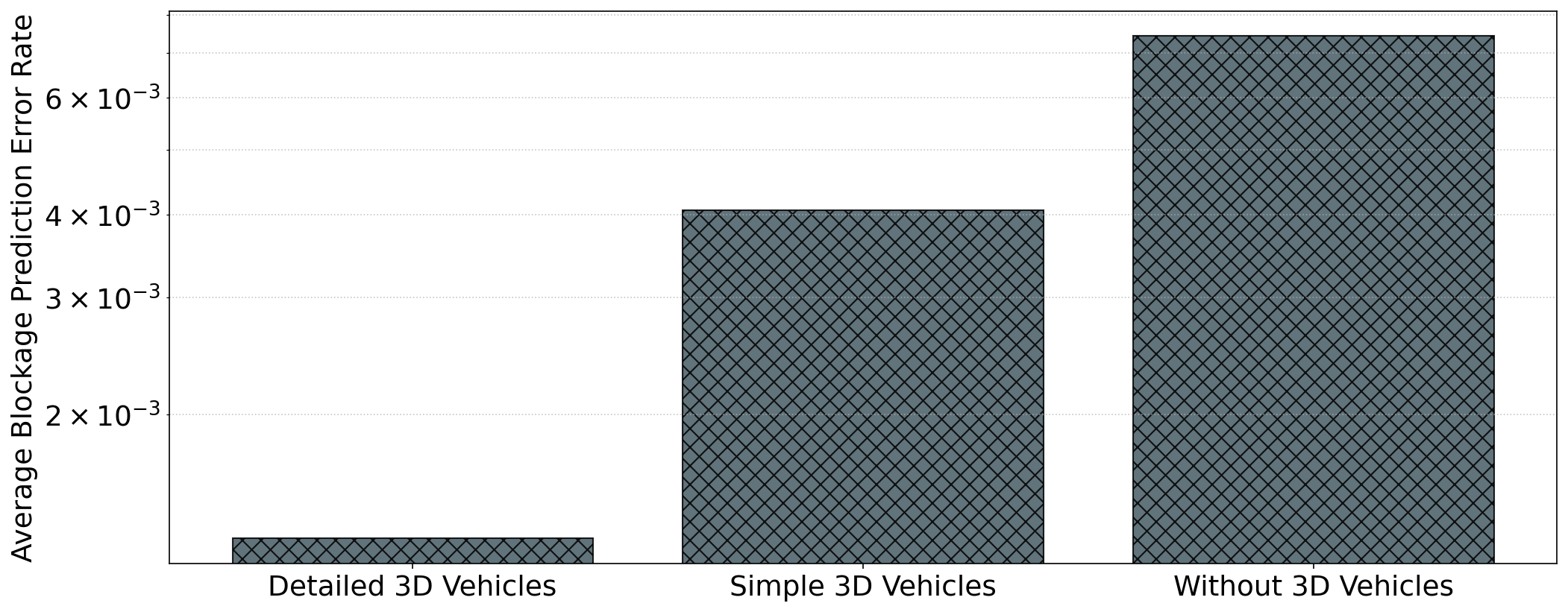}
        \label{fig:3D_Blockage}
    }
    
    \subfloat[Latency]{
        \includegraphics[width=0.78\columnwidth]{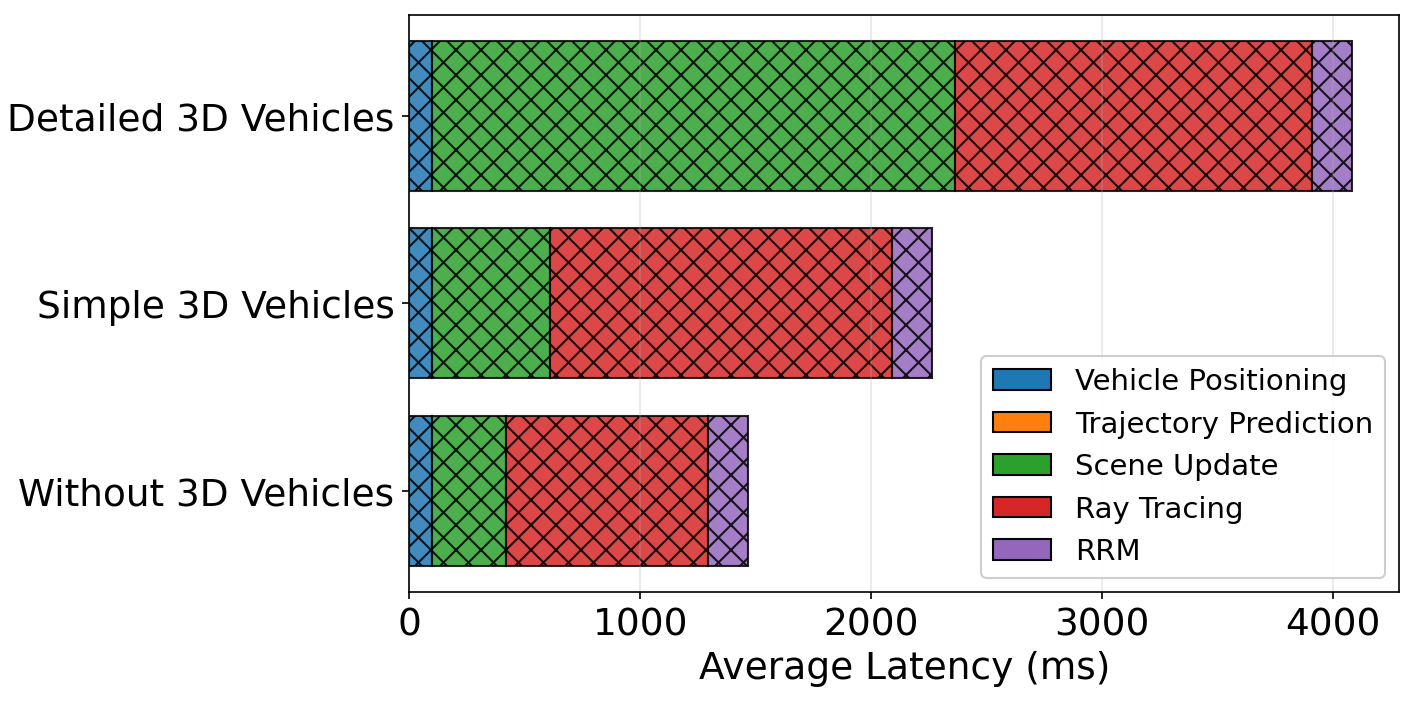}
        \label{fig:3D_Latency}
    }
    \caption{Impact of 3D modeling of vehicles in terms of blockage prediction error rate and latency for three versions of the high-fidelity reference NDT; with detailed 3D models of vehicles, with simplified 3D models of vehicles, and without 3D models of vehicles, for high-density Ottawa scenario.}
    \label{fig:3D}
\end{figure}

\textit{Blockage prediction error rate.} A link $(u,b)$ at distance $d_{u,b}$ is declared LOS, $\mathcal{I}_{u,b}(C) = 1$, if its path gain $G_{u,b}^{C}$ lies within $\Delta = 15$\,dB of the free-space bound $N_{tx}\big(c/(4\pi f_c d_{u,b})\big)^2$. This margin exceeds the pattern and material-loss spread of unobstructed links but is below the excess loss of a vehicle body. The error rate is the fraction of links whose LOS state differs from that of the reference,
\begin{equation}
P_{\mathrm{blk}} = \frac{1}{|\mathcal{U}||\mathcal{B}|}\sum_{u,b} \mathbf{1}\big\{\mathcal{I}_{u,b}(C) \neq \mathcal{I}_{u,b}(C_{GT})\big\}. \label{eq:blk}
\end{equation}
To evaluate the impact of 3D modeling of vehicles on channel prediction, we compare three versions of the high-fidelity reference NDT: with detailed 3D models of vehicles, with simplified 3D models of vehicles, and without 3D models of vehicles. The evaluated metrics are blockage prediction error rate and latency. The results are shown in Fig.$\,$\ref{fig:3D}, demonstrating that omitting 3D models of vehicles increases the blockage prediction error rate, as the simulation fails to capture the severe line-of-sight (LOS) obstructions caused by dynamic traffic, specifically large vehicles like buses and trucks. Conversely, utilizing detailed 3D vehicles achieves the lowest blockage prediction error rate but incurs prohibitive computational latency for scene updating as the vehicle moves, rendering it unsuitable for real-time edge execution. The proposed approach of using simplified low-poly 3D vehicle models strikes a favorable balance: it accurately predicts macroscopic blockages at a much lower scene update latency. This ablation also covers vehicle-mesh mismatch; a more detailed analysis of vehicle-induced blockage is given in our prior work \cite{Makvandi2025Offline, Makvandi2026AIRTwin}.
\setlength{\textfloatsep}{0pt}


\subsubsection{\textbf{Trajectory Prediction, Continual Learning, and Positioning Error}}
\begin{figure}[!t]
    \centering
    \subfloat[FDE]{
        \includegraphics[width=\columnwidth]{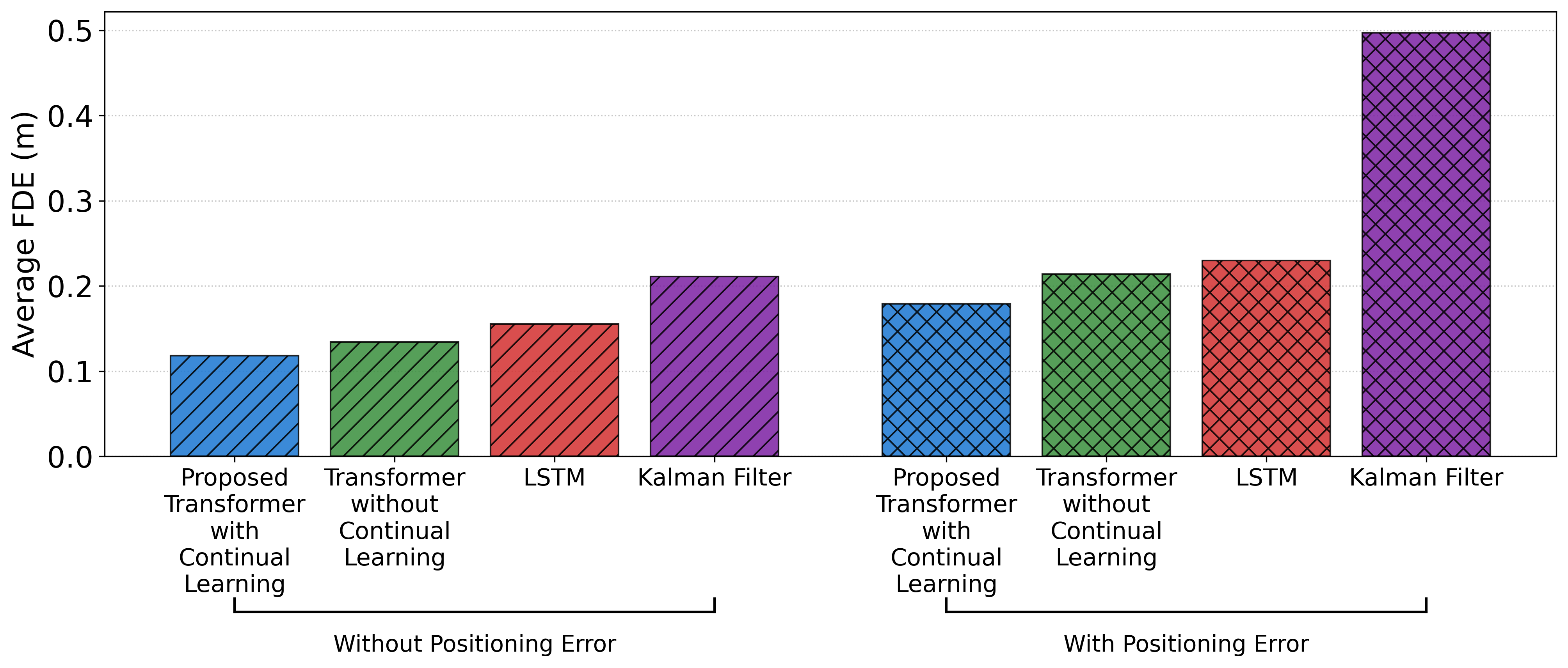}
        \label{fig:VTP_FDE}
    }
    
    \subfloat[Path gain RMSE]{
        \includegraphics[width=0.81\columnwidth]{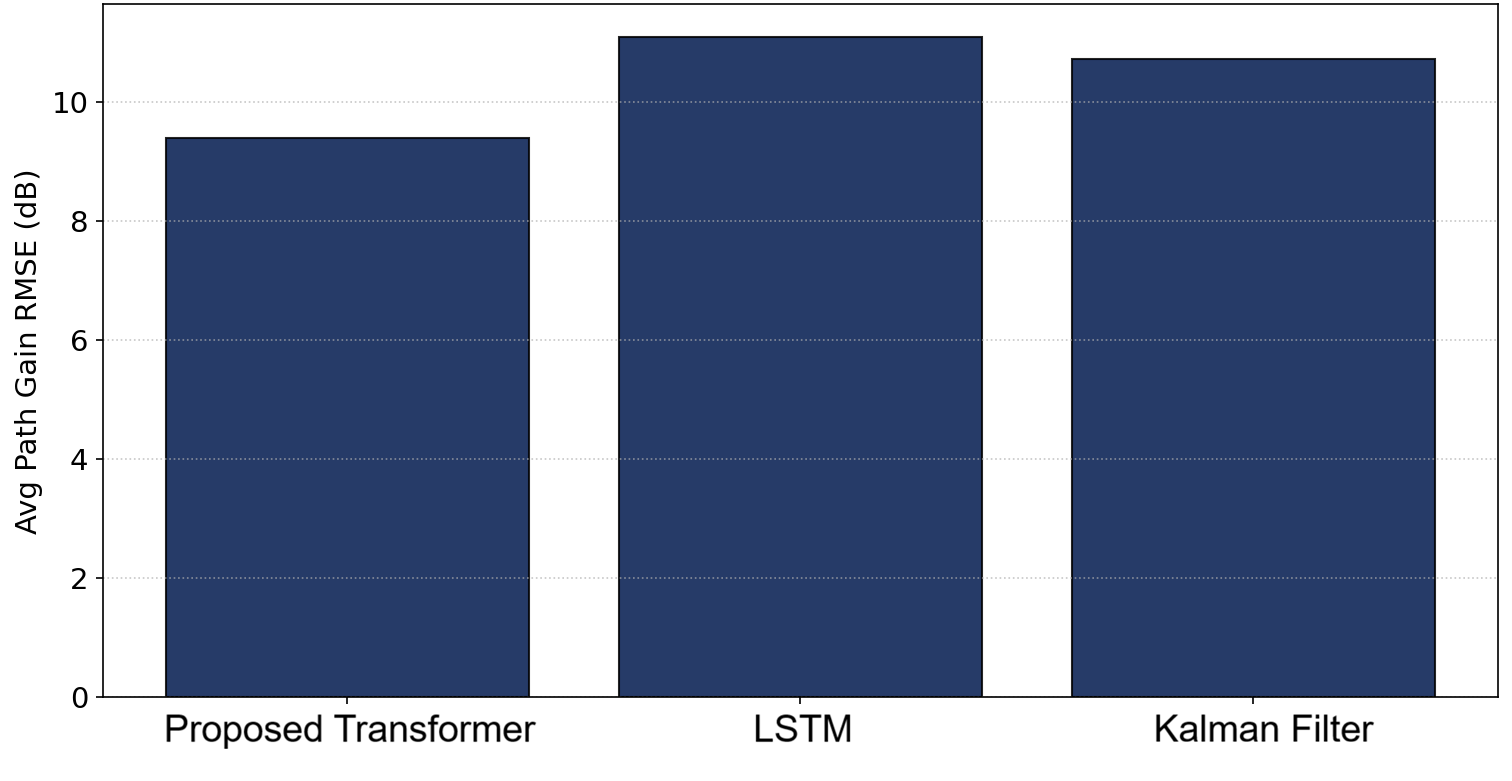}
        \label{fig:VTP_RMSE}
    }
    \caption{Comparison of vehicle trajectory prediction models in terms of (a) FDE without and with positioning error, including the proposed Transformer without continual learning,} and (b) path gain RMSE.
    \label{fig:VTP}
\end{figure}

The accuracy of the underlying trajectory prediction engine directly dictates the reliability of the predictive AdaPTwin. We compare the proposed Transformer with Continual Learning against an LSTM-based predictor, and a kinematic Constant Velocity model augmented with a Kalman Filter (KF). Figure $\,$\ref{fig:VTP_FDE} shows the comparison of these models in terms of FDE, without and with positioning error, using a real-world dataset in an urban area in Chongqing, China \cite{Xu2022SinD}\footnote{The neural predictors are trained on the Tianjin recordings of the SinD dataset \cite{Xu2022SinD}, and all predictors are tested on its Chongqing recordings, i.e., on an intersection not seen in training, to stress their generalization.}. Figure $\,$\ref{fig:VTP_RMSE} shows the comparison of three versions of the predictive NDT utilizing different trajectory prediction models in terms of path gain RMSE using the SUMO-generated Vancouver dataset.
\setlength{\textfloatsep}{0pt} 
As illustrated, the constant velocity with Kalman Filter baseline fails to capture non-linear movements at complex intersections, resulting in the highest FDE. While the LSTM model captures basic temporal dynamics, it struggles with long-term dependencies in dense traffic. The predictive NDT with the KF nevertheless achieves a lower path gain RMSE than with the LSTM, because the channel depends not only on the positions of the vehicles but also on their headings, which the KF predicts better. The proposed Transformer, by leveraging a self-attention mechanism to capture long-term dependencies and continual learning to adapt to new environments and mobility patterns, achieves the lowest FDE and subsequently the lowest RMSE. Fig.~\ref{fig:VTP_FDE} also serves as an ablation of continual learning: freezing the offline-trained weights raises the FDE of the Transformer from $0.119$ to $0.135$\,m. To relax the GNSS-RTK assumption, zero-mean Gaussian errors with a standard deviation of $5$\,cm per coordinate, several times the typical RTK error, are added to all observed positions, and the FDE is computed against the true positions. The FDE of all predictors increases, but the proposed Transformer remains the most accurate ($0.180$\,m) and shows the smallest increase ($0.061$\,m). At a typical distance of $100$\,m, an error of $0.18$\,m subtends about $0.1^\circ$, well within the $8^\circ$ spacing of the $16$-beam codebook, so its impact on beam selection is expected to be small. The same noise model covers position reports corrupted by an imperfect feedback channel. The delay of the position feedback is part of $T_P$ and is charged to the latency budget \eqref{eq:budget}. An unknown clock offset $\delta$ between the vehicles and the twin, in contrast, causes a systematic rather than random error: as the inputs are ego-centric, the history of a vehicle moving at a nearly constant speed $v$ is merely shifted, and its prediction is biased by about $v\delta$ along its heading, which raises the FDE by at most about $v\delta$.

\subsubsection{\textbf{Edge-to-Cloud Latency Ratio}}
As both tiers ran on one GPU, a slower edge is emulated by scaling the cloud-measured latencies of ray tracing, scene update, and RRM by $\kappa$ (Vancouver, medium density); $T_P$ and $T_{TP}$ are edge-side quantities and are not scaled. Fig.~8  compares the end-to-end latency estimated by Phase~1 and the sum rate of the proposed $\kappa$-aware selection with those of a $\kappa$-oblivious policy that keeps the fidelity configuration selected at $\kappa = 1$. As $\kappa$ grows, $T_{Budget}$ in \eqref{eq:budget} and the set of feasible configurations shrink, and Phase~1 moves to leaner configurations; when no configuration is feasible, the leanest one is used (Algorithm~\ref{alg:fidelity}). Hence, the latency of the $\kappa$-aware selection remains below the deadline, and its sum rate remains nearly constant, decreasing only slightly when the leanest configuration is used. Although the all-link path gain RMSE of the leaner configurations increases, the sum rate is less sensitive to it, since this RMSE also includes deeply shadowed links that are not serving links. In contrast, the latency of the $\kappa$-oblivious policy grows with $\kappa$ and exceeds the deadline, so that its sum rate decreases steadily under the deadline rule.

\setlength{\textfloatsep}{0pt}
\begin{figure}[!t]
    \centering
    \subfloat[Latency]{
        \includegraphics[width=0.75\columnwidth]{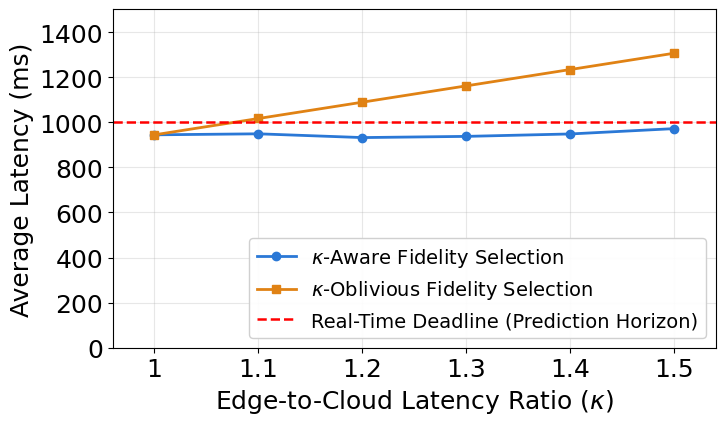}
        \label{fig:Kappa_Latency}
    }

    \subfloat[Sum Rate]{
        \includegraphics[width=0.75\columnwidth]{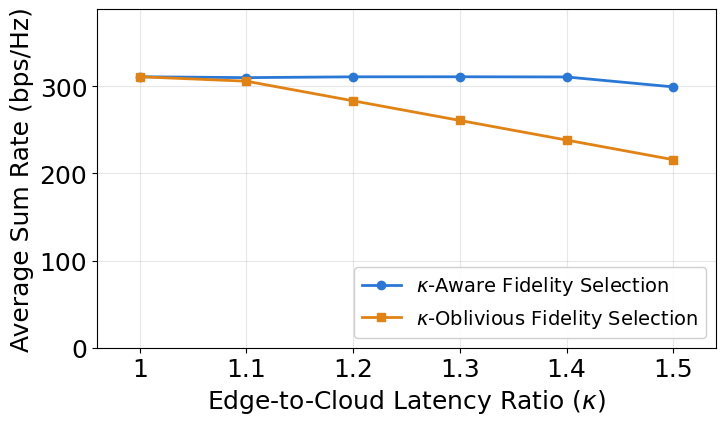}
        \label{fig:Kappa_Rate}
    }
    \caption{Latency and sum rate of $\kappa$-aware and $\kappa$-oblivious fidelity selection versus $\kappa$ (Vancouver, medium density).}
    \label{fig:Kappa}
\end{figure}

\vspace{-0.3cm}
\subsection{Optimality and Scalability}
\label{sec:optimality}
\subsubsection{\textbf{Proposed Heuristic RRM}}
\begin{figure}[H]
    \centering
    \subfloat[Sum Rate]{
        \includegraphics[width=0.8\columnwidth]{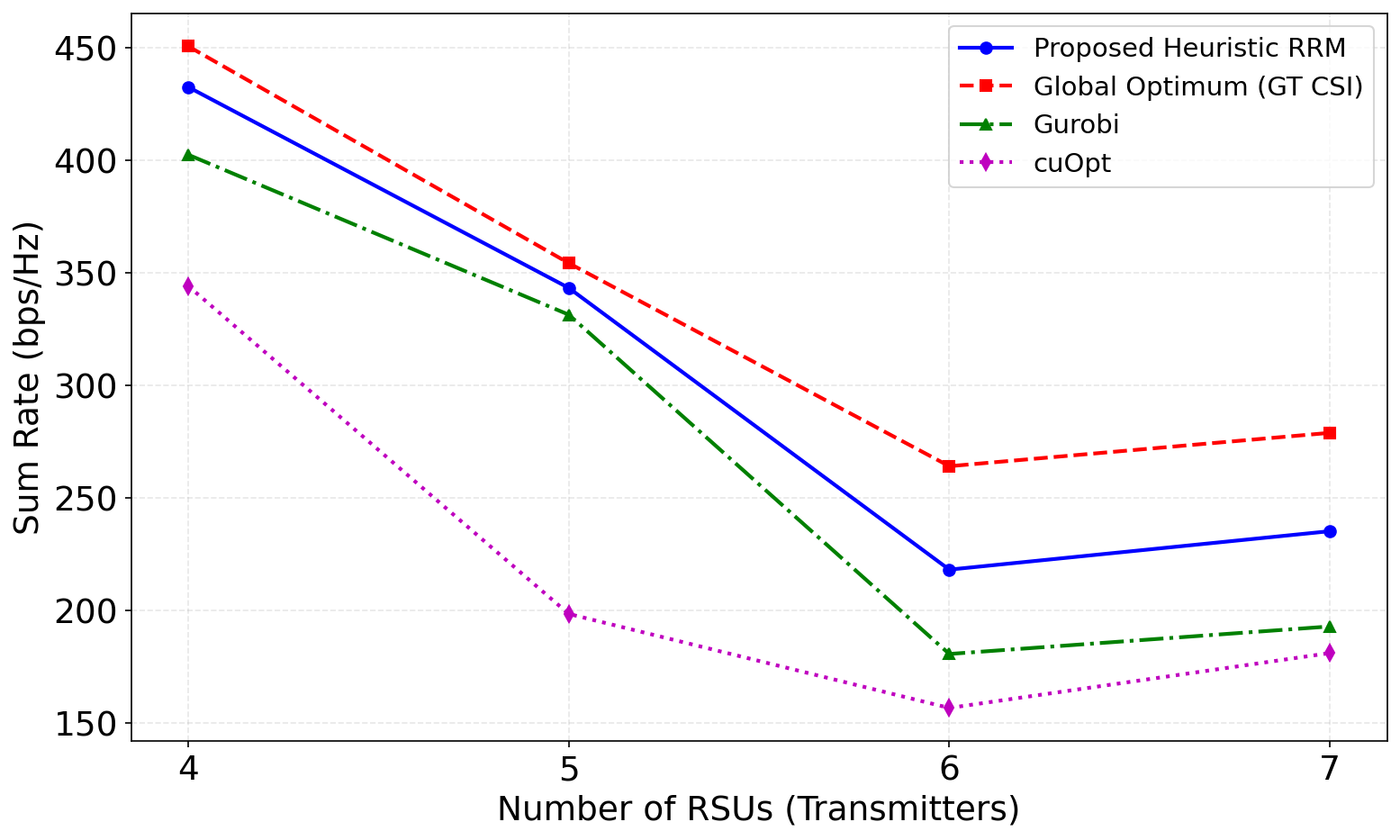}
        \label{fig:RRM_Rate}
    }
    
    \subfloat[RRM Latency]{
        \includegraphics[width=0.8\columnwidth]{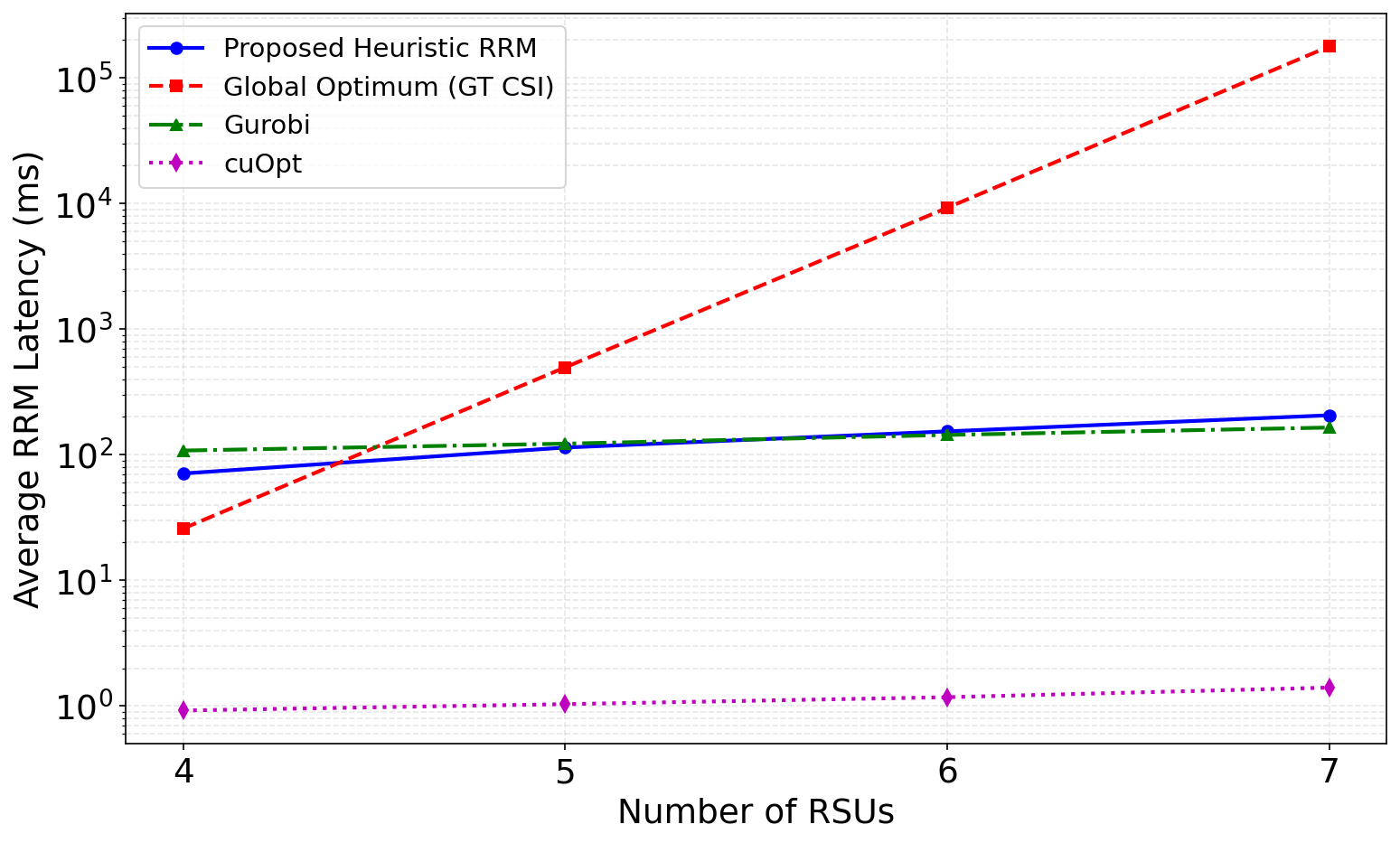}
        \label{fig:RRM_Latency}
    }
    \caption{Comparison of the proposed heuristic RRM with global optimum, Gurobi, and cuOPT solvers in terms of sum rate and latency.}
    \label{fig:RRM}
\end{figure}

We first assess the performance and scalability of the proposed Multi-Start Iterative Coordinate Descent RRM algorithm. It is benchmarked against the Global Optimum (exhaustive search), as well as general-purpose heuristic solvers (Gurobi and NVIDIA cuOPT). The metrics considered are sum rate, computational latency, and scalability with an increasing number of RSUs. Figure $\,$\ref{fig:RRM_Rate} and $\,$\ref{fig:RRM_Latency} show the comparison of these approaches in terms of sum rate and latency, respectively, for different number of RSUs in the high density scenario in Ottawa. The results show that the Global Optimum provides the theoretical upper bound for the sum rate but exhibits an exponential growth in execution time $\mathcal{O}(W^B)$, as number of RSUs increases, and thus, it exceeds the real-time latency budget. The proposed heuristic RRM algorithm satisfies the latency constraints for different numbers of RSUs, confirming its scalability, while achieving a higher sum rate compared to Gurobi and cuOPT solvers. Its combined loss with the two-timescale decomposition is quantified next.

\subsubsection{\textbf{Joint Optimization vs. Two-Timescale Decomposition}}
\label{sec:joint}

\begin{figure*}[!t]
    \centering
    \subfloat[Latency]{
        \includegraphics[width=0.27\linewidth]{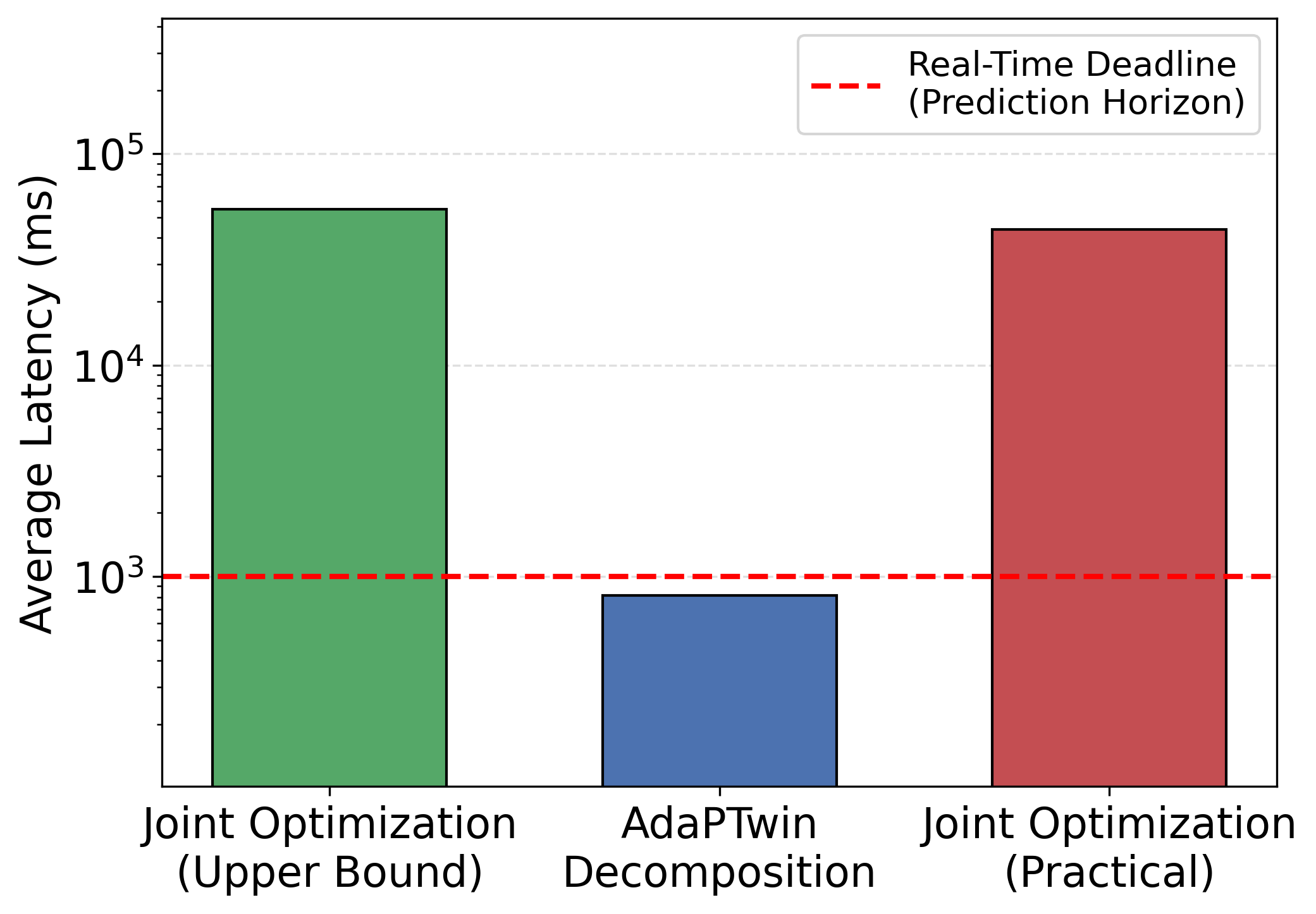}
        \label{fig:Joint_Latency}
    }
    \hfill
    \subfloat[Sum Rate]{
        \includegraphics[width=0.29\linewidth]{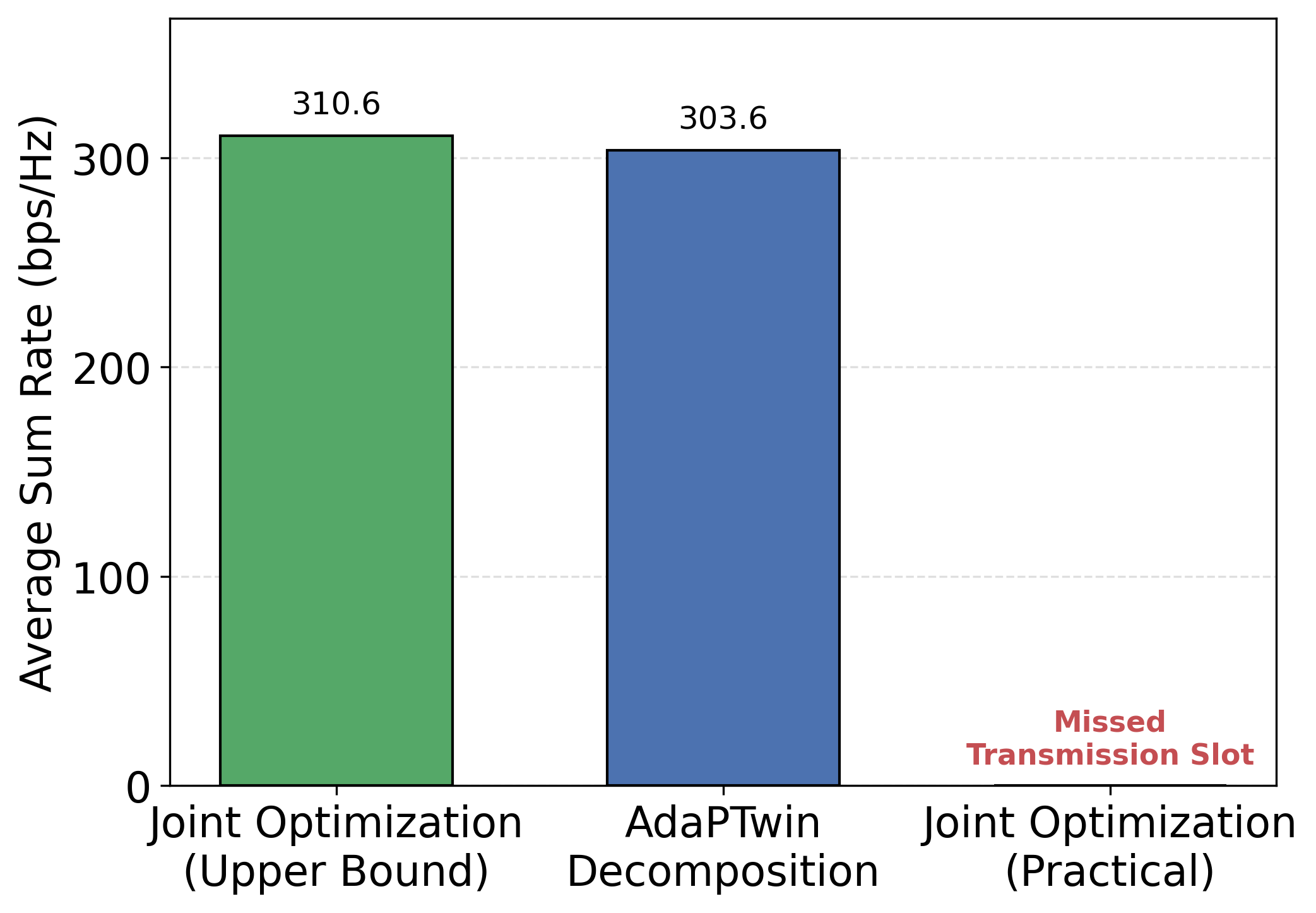}
        \label{fig:Joint_Rate}
    }
    \hfill
    \subfloat[Outage Probability]{
        \includegraphics[width=0.29\linewidth]{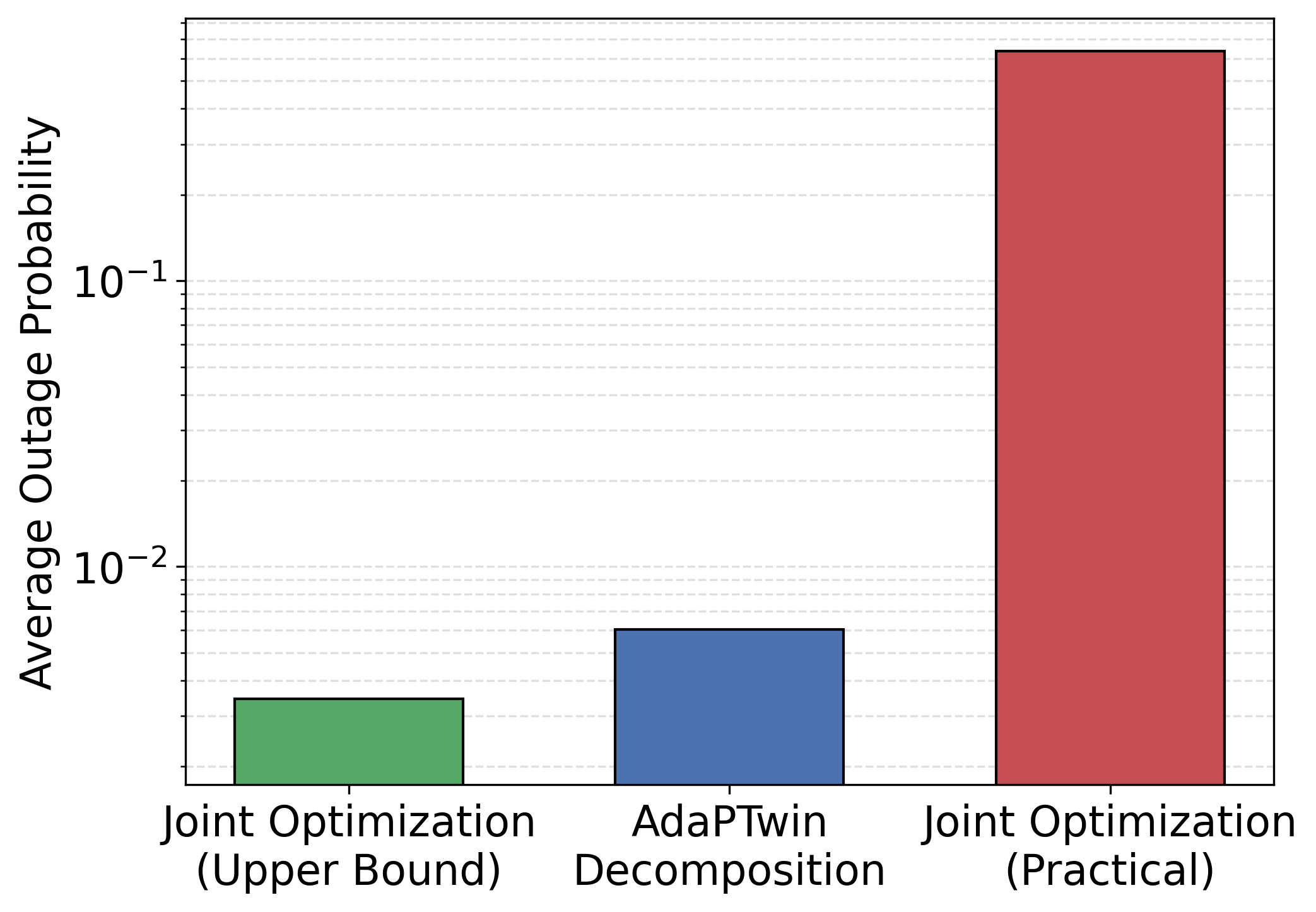}
        \label{fig:Joint_Outage}
    }
    \caption{AdaPTwin versus per-slot joint optimization (Vancouver, medium density): upper bound with exhaustive-search RRM, and practical joint optimization with Algorithm~\ref{alg:rrm} applied after its own latency. (a) Latency, (b) sum rate, and (c) outage probability.}
    \label{fig:Joint}
\end{figure*}

To quantify the loss of the decomposition, we solve $\mathcal{P}_{\text{joint}}$ per slot at $9$ snapshots of the evaluation window (Vancouver, medium density) and report averages over these snapshots. For the predicted positions, each configuration of a candidate grid is ray-traced, and the latency-feasible configuration with the highest proportional-fairness score on the reference channels (obtained with oracle knowledge) is selected. The grid comprises $60$ configurations of $\mathcal{C}_{edge}$ ($F_V \in \{30, 50\}$, $D_{max} \in \{1, 3, 5, 7, 9, 10\}$, and $N_{rays} \in \{10^2, 10^3, \dots, 10^6\}$), which span the full range of every fidelity parameter of $\mathcal{C}_{edge}$ and include AdaPTwin's configuration. Unlike Phase~1, which runs once per $T_{\text{update}}$, the joint search must ray-trace every candidate and solve the RRM problem for each of them in every slot, so a denser grid would only further increase its latency. The \textit{upper bound} solves the RRM problem by an exhaustive search over all $16^5$ beam states and evaluates the resultant decisions at the instant for which they were computed, ignoring the computation time. The \textit{practical} joint optimization uses Algorithm~3 instead, which makes it much faster, and its decision is evaluated at the instant at which it becomes available, i.e., after its computation time. Both are evaluated on the reference channels. As shown in Fig.~10, AdaPTwin attains $303.6$\,bps/Hz against $310.6$\,bps/Hz for the upper bound, a loss of only $2.3\%$, with an outage probability of $0.006$, whereas the outage of the upper bound remains further below. Before its latency is accounted for, the practical joint optimization also reaches $310.6$\,bps/Hz. Therefore, the sum rate loss of AdaPTwin with respect to the per-slot joint optimization mainly stems from the fidelity selection which is made once per $T_{\text{update}}$ via $\widehat{\mathcal{E}}(C_k)$. 
However, the joint search takes $54.6$\,s per decision with exhaustive search and still $43.9$\,s with Algorithm~3, against $0.81$\,s for AdaPTwin. Therefore, in practice, the decision of per-slot joint optimization becomes available about $44$\,s after the instant for which it was computed, and it contains an association only for the vehicles present at that instant. When it is applied,  many of the vehicles then present were not present at the instant for which it was computed and remain unassociated, and the others are served with beams selected for their earlier positions. Hence, the outage probability of the  joint search in \textit{practice}  rises to $0.64$, against $0.01$ for AdaPTwin at the same instants. Under the deadline rule, its sum rate is zero. Moreover, re-selecting the NDT fidelity configuration for $\lambda \in \{0, 0.5, 1, 2, 5, 10, 20\}$ changes the sum rate of the upper bound by at most $1.5\%$.

\vspace{-0.3cm}
\subsection{Statistical Reliability}
\label{sec:mc}
\begin{figure}[!t]
    \centering
    \includegraphics[width=0.75\columnwidth]{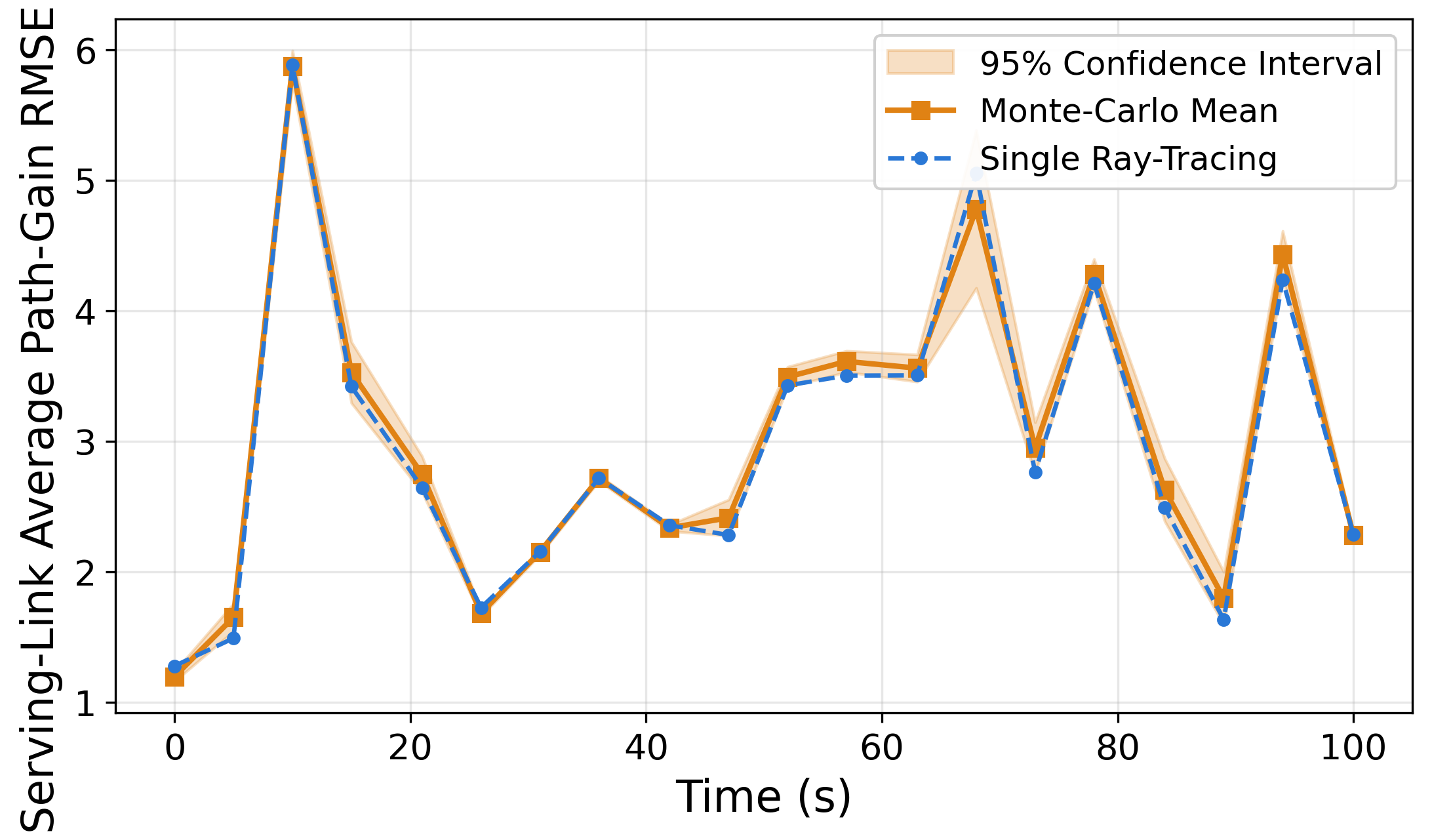}
    \caption{Serving-link path gain RMSE (dB) of a single realization (``Single Ray-Tracing'') vs. the Monte-Carlo mean and 95\% confidence interval of the mean over $10$ seeds (Vancouver, medium density).}
    \label{fig:MC}
\end{figure}

To verify that the results are not artifacts of a single random realization of ray tracing and RRM, we repeat the AdaPTwin pipeline with $10$ seeds at $20$ snapshots (Vancouver, medium density). As shown in Fig.~\ref{fig:MC}, the serving-link path gain RMSE of the single realization used in this paper closely follows the Monte-Carlo mean at every snapshot: on average, it is $2.95$\,dB against a Monte-Carlo mean of $3.00$\,dB ($95\%$ confidence interval of the mean: $\pm 0.04$\,dB), a deviation of $1.7\%$ that is within one standard deviation across seeds, and the resulting sum rate deviates by only $0.13\%$. Over all $101$ snapshots, AdaPTwin attains $311.3 \pm 6.3$\,bps/Hz and a latency of $788.5 \pm 7.6$\,ms (mean $\pm$ half-width of the $95\%$ confidence interval). It is lower than the latency estimated by Phase~1 (Fig.~8), since Phase~1 evaluates the instant with the largest number of vehicles.

\setlength{\textfloatsep}{0pt}

\vspace{-0.3cm}
\section{Discussion: Scope and Limitations}
\textit{Decision timescale.} At $5.875$\,GHz and $50$\,km/h, the coherence time is about $1.6$\,ms, so a predicted CIR does not track small-scale fading over the $1$\,s horizon; it provides the large-scale channel on which the semi-persistent beam and association decisions depend. AdaPTwin is thus a proactive RRM framework at the beam-management timescale, not a URLLC link-level solution. Within a slot, a vehicle at $50$\,km/h moves about $14$\,m, i.e., roughly one $8^\circ$ beam spacing at $100$\,m; since the rates are evaluated at the predicted instant, shorter slots would reduce this intra-slot mismatch. Since the non-ray-tracing latencies and the guard time sum to about $720$\,ms at $\kappa = 1$, shorter horizons require faster positioning, scene updates, and RRM.

\textit{Validation scope.} The computation latencies are measured on a single GPU, and the trajectory predictor is also evaluated on real-world recordings (Section~VII-E.2). The channel accuracy, in contrast, is relative to a high-fidelity reference NDT and excludes errors in scene reconstruction and materials; since all NDTs share the scene and the reference, the comparisons isolate the effect of the fidelity configuration and the decision latency. Ray tracing with calibrated geometry and materials reproduces measured urban propagation well \cite{DegliEsposti2014RT, Hoydis2024LearningRadio}, and replacing the reference by measured channels would change $\widehat{\mathcal{E}}(C_k)$ in \eqref{eq:err} but neither Algorithm~2 nor the rest of the framework. A measurement-anchored version would calibrate the materials by differentiable ray tracing against channel soundings and measure $\kappa$ on a real MEC node. Such a campaign, as well as generalization beyond two dense urban areas and a downlink single-band setting, is outside the scope of this paper.

\vspace{-0.3cm}

\section{Conclusion}
This paper introduced AdaPTwin, an adaptive multi-fidelity predictive NDT designed for proactive and situation-aware RRM in highly dynamic vehicular networks. AdaPTwin predicts the vehicle--RSU channels ahead of time by forecasting vehicle trajectories with a Transformer enhanced by continual learning and by ray-tracing the predicted 3D scene of buildings, roads, and vehicles with NVIDIA Sionna. To maximize channel prediction accuracy under strict latency constraints, we develop a hierarchical cloud–edge architecture that leverages cloud-based processing for periodic, computationally intensive digital twin calibration, while enabling real-time trajectory prediction, ray tracing, and RRM at the edge. By leveraging this NDT-enabled proactive RRM pipeline, a joint  RSU beamforming and vehicle-RSU association problem is solved using a scalable multi-start iterative coordinate descent heuristic algorithm, with the objective of maximizing proportional fair network sum-rate. Simulation results show that, unlike existing reactive, non-adaptive, and single-fidelity NDT-based RRM approaches, AdaPTwin adapts to the evaluated traffic densities and urban environments: it achieves up to a 70\% increase in sum rate and an 80\% reduction in outage probability compared with a non-adaptive predictive multi-fidelity NDT while meeting the per-slot deadline. It also remains within $2.3\%$ of a per-slot joint-optimization upper bound and keeps meeting the deadline on an emulated edge slower than the cloud. For future work, we will calibrate the twin against channel measurements and consider the uncertainty and confidence level of predictions for improving the reliability of AdaPTwin for vehicular networks. Furthermore, we will consider safety awareness in RRM with the aim of jointly improving the safety of autonomous vehicles and network efficiency.

\vspace{-0.3cm}

\bibliographystyle{IEEEtran}
\bibliography{Bibliography}

\balance
\end{document}